\documentclass[a4paper,UKenglish,cleveref, autoref, thm-restate]{lipics-v2021}

\title{Shared-Memory $n$-level Hypergraph Partitioning} 

\titlerunning{Shared-Memory $n$-level Hypergraph Partitioning} 

\author{Lars Gottesbüren}{Karlsruhe Institute of Technology, Karlsruhe, Germany}{lars.gottesbueren@kit.edu}{}{}

\author{Tobias Heuer}{Karlsruhe Institute of Technology, Karlsruhe, Germany}{tobias.heuer@kit.edu}{}{}

\author{Peter Sanders}{Karlsruhe Institute of Technology, Karlsruhe, Germany}{sanders@kit.edu}{}{}

\author{Sebastian Schlag}{Karlsruhe Institute of Technology, Karlsruhe, Germany}{research@sebastianschlag.de}{}{}

\authorrunning{L. Gottesbüren, T. Heuer, P. Sanders, S. Schlag} 

\Copyright{Lars Gottesbüren, Tobias Heuer, Peter Sanders, Sebastian Schlag} 



\category{} 

\relatedversion{} 

\supplement{The source code and data have been made available at \url{https://github.com/kahypar/mt-kahypar} and \url{http://algo2.iti.kit.edu/heuer/nlevel/}}

\acknowledgements{This work was partially supported by DFG grants WA654/19-2, SA933/10-2 and SA933/11-1.
The authors acknowledge support by the state of Baden-W\"urttemberg through bwHPC.}

\nolinenumbers 

\hideLIPIcs  

\EventEditors{John Q. Open and Joan R. Access}
\EventNoEds{2}
\EventLongTitle{42nd Conference on Very Important Topics (CVIT 2016)}
\EventShortTitle{CVIT 2016}
\EventAcronym{CVIT}
\EventYear{2016}
\EventDate{December 24--27, 2016}
\EventLocation{Little Whinging, United Kingdom}
\EventLogo{}
\SeriesVolume{42}
\ArticleNo{23}

\usepackage[algo2e,ruled,vlined,linesnumbered, noend]{algorithm2e}
\usepackage[utf8]{inputenc}
\usepackage[TS1,T1]{fontenc}
\SetCommentSty{textit}
\DontPrintSemicolon
\IncMargin{-\parindent}
\SetAlCapHSkip{0pt}
\SetAlgoLined

\SetKwFor{ParallelFor}{for}{do in parallel}{}

\usepackage{latexsym,amsmath,textcomp,pifont,marvosym}
\usepackage{wasysym}
\usepackage{tikz}
\usetikzlibrary{calc}

\usepackage{multirow}
\usepackage{bm}
\usepackage{datatool}
\usepackage{booktabs}
\DTLsetseparator{ = }

\usepackage{pgfplots}
\usepgfplotslibrary{external}
\tikzsetexternalprefix{experiments/}

\usepackage{rotating}

\usepackage{mathtools}

\newcommand{\placeholder}[2]{\DTLfetch{#1}{key}{#2}{value}}
\newcommand{\externalizedfigure}[2]{
  \includegraphics{experiments/mt_kahypar-figure#1}
}

\newcommand{\splitatcommas}[1]{%
  \begingroup
  \begingroup\lccode`~=`, \lowercase{\endgroup
    \edef~{\mathchar\the\mathcode`, \penalty0 \noexpand\hspace{0pt plus 1em}}%
  }\mathcode`,="8000 #1%
  \endgroup
}

\newcommand{\mtkahypar}{\texttt{Mt-KaHyPar}}

\newcommand{\kahyparhfc}{\texttt{KaHyPar-HFC}~\cite{KAHYPAR-HFC}}

\newcommand{\mtkahyparold}{Mt-KaHyPar-D}
\newcommand{\mtkahyparnew}{Mt-KaHyPar-Q}

\newcommand{\Oh}[1]{\ensuremath{\mathcal{O}(#1)}}

\newcommand{\Partition}{\ensuremath{\mathrm{\Pi}}}%
\newcommand{\incnets}{\ensuremath{\mathrm{I}}}%
\newcommand{\pinsinpart}{\ensuremath{\mathrm{\Phi}}}

\newcommand{\con}{\ensuremath{\lambda}}
\newcommand{\conset}{\ensuremath{\Lambda}}

\newcommand{\maxsize}[1]{\ensuremath{\Delta_{#1}}}

\newcommand{\meddeg}{\ensuremath{\widetilde{d(v)}}}
\newcommand{\medsize}{\ensuremath{\widetilde{|e|}}}

\newcommand{\Partitioner}[1]{\textsf{#1}} 
\newcommand{\SeqAlgo}[1]{\Partitioner{#1}}
\newcommand{\ExpAlgo}[2]{\Partitioner{#1~#2}}

\definecolor{fuchsiapink}{rgb}{1.0, 0.47, 1.0}

\definecolor{utahcrimson}{rgb}{0.83, 0.0, 0.25}

\definecolor{nothappycolor}{rgb}{0.5, 0.5, 0.5}

\newcommand{\plusplus}{\texttt{++}}

\newcommand{\gpp}[1]{g\plusplus#1}

\newcommand{\doublelinkedlist}[1]{\ensuremath{L_{#1}}}

\newcommand{\pending}{\ensuremath{\text{\textsf{pending}}}}
\newcommand{\rep}{\ensuremath{\text{\textsf{rep}}}}

\newcommand{\contractions}{\ensuremath{\mathcal{C}}}
\newcommand{\contractionforest}{\ensuremath{\mathcal{F}}}
\newcommand{\maxbatchsize}{\ensuremath{b_{\max}}}
\newcommand{\batches}{\ensuremath{\mathcal{B}}}
\newcommand{\currentbatch}{\ensuremath{B_{\text{cur}}}}

\begin{document}

\maketitle

\begin{abstract}
  We present a shared-memory algorithm to compute high-quality solutions to the balanced $k$-way hypergraph partitioning problem.
  This problem asks for a partition of the vertex set into $k$ disjoint blocks of bounded size that minimizes the connectivity metric (i.e., the sum of the number of different blocks connected by each hyperedge).
  High solution quality is achieved by parallelizing the core technique of the currently best
  sequential partitioner KaHyPar: the most extreme $n$-level version
  of the widely used multilevel paradigm, where only a single vertex is contracted on each level.
  This approach is made \emph{fast} and \emph{scalable} through
  intrusive algorithms and data structures that allow precise control of parallelism through atomic operations
  and fine-grained locking.
  We perform extensive experiments on more than 500 real-world hypergraphs with up to $140$ million vertices and two billion
  pins (sum of hyperedge sizes). We find that our algorithm computes solutions that are on par with a comparable configuration of KaHyPar
  while being an order of magnitude faster on average. Moreover, we show that recent non-multilevel
  algorithms specifically designed to partition large instances have considerable quality penalties and no
  clear advantage in running time.
\end{abstract}

\section{Introduction}

Hypergraphs are a generalization of graphs where each hyperedge can connect more than two vertices.
Processing complex hypergraphs has become an important application area but also a major challenge~\cite{HeintzC14}, in particular in parallel scenarios.
In this context, hypergraph partitioning (HGP) is a crucial operation.
It asks for partitioning the vertex set into $k$ disjoint \emph{blocks} of about the same size while minimizing the number of cut edges (often additionally weighted by the number of different blocks connected by cut hyperedges).
The problem is again formally defined in \cref{s:preliminaries}.
HGP has applications in various domains such as
VLSI design~\cite{ALPERT-SURVEY}, scientific computing~\cite{PATOH},
hypergraph processing frameworks~\cite{hg-processing-framework-MESH, hg-processing-hyperx},
and storage sharding in distributed database management
systems~\cite{schism, clay, hepart, sword}.

Since the HGP problem is NP-hard~\cite{LENGAUER} and even hard to
approximate~\cite{BUI}, heuristic algorithms are used in practice.
In particular, \emph{multilevel} algorithms~\cite{KAHYPAR-K, PATOH, HMETIS, MONDRIAAN}
have emerged as an excellent compromise between speed and quality.
Multilevel HGP proceeds in three phases: First, the hypergraph is \emph{coarsened} to obtain a
hierarchy of structurally similar and successively smaller hypergraphs
by \emph{contracting} pairs or clusters of vertices. We can now afford
to apply expensive \emph{initial partitioning} algorithms to the
resulting small hypergraph that still represents essential features of
the input.
Finally, coarsening is successively undone.
At each level of the hierarchy, vertices are assigned to the same block of the partition as their constituent on the previous level, before \emph{refinement} algorithms are used to improve the quality of the solution.
This allows to perform optimization on many levels of granularity using simple local operations.

Until recently, sequential algorithms were dominant since existing distributed systems~\cite{PARKWAY-2,ZOLTAN,SHP} sacrifice solution quality by using comparatively weak components that are easier to parallelize.
Two sequential systems stand out.
\Partitioner{PaToH} \cite{PATOH} is very fast and computes partitions of reasonable quality, whereas \Partitioner{KaHyPar}~\cite{KAHYPAR-CA,KAHYPAR-DIS} computes high-quality partitions, due to its $n$-level approach~\cite{ADAPTIVE-STOP-RULE}.
It uses a dynamic hypergraph data structure in order to support a maximally fine-grained hierarchy -- in each step of coarsening, only a single pair of vertices is contracted.
Correspondingly, in each step of refinement, only a single vertex is uncontracted, allowing a highly localized search for improvements.

In a previous paper \cite{MT-KAHYPAR}, we presented a multi-threaded high-performance HGP system that uses a traditional multilevel approach with few levels.
Even with few cores, this system considerably outperforms \Partitioner{PaToH} as well as most of the sequential and all parallel HGP codes we are aware of, both in terms of solution quality and running time.
\Partitioner{KaHyPar}~\cite{KAHYPAR-CA} still computes better partitions due to its n-level approach but is substantially slower -- prohibitively so for large hypergraphs.

\subparagraph*{Contributions and Outline.}

In this paper, we address the challenge of achieving the quality of \Partitioner{KaHyPar} in a fast parallel code.
Our approach is to parallelize the $n$-level variant even though it looks inherently sequential and even though it requires complex dynamic data structures.
To this end, we develop a representation of contractions by a forest and an approach to decompose the forest into batches of local improvements that can be performed in parallel.
This is aided by a space-efficient, concurrent and dynamic hypergraph data structure that is used in a highly intrusive fashion.
For example, we exploit internal states observed during concurrent uncontraction operations for updating entries in data structures used by the refinement algorithms.
We also develop an adaptive selection scheme for flat initial bipartitioning algorithms.
In combination with numerous implementation level improvements this also accelerated our previous system~\cite{MT-KAHYPAR} by a factor of about 1.5.


We support our findings with extensive experiments on orders of magnitude more instances than many previous, high-impact papers.
Since the speed and quality of HGP systems varies widely from input to input, this helps to clarify the relative merits of different approaches.
For example, we also consider two non-multilevel systems~\cite{SHP, TheHypeIsOver} that have recently been considered and show that they have enormous quality penalties compared to multilevel systems while not giving consistent performance advantages over the fastest multilevel codes.

In \cref{s:preliminaries}, we introduce notation, used concepts and formally define the problem, before discussing related work in \cref{s:relatedWork}.
Then \cref{s:main} presents the algorithmic components of our new system \Partitioner{\mtkahyparnew}, though we mostly focus on parallel coarsening and uncoarsening.
Experiments in \cref{s:experiments} indicate that our system achieves similar quality as a comparable variant of \Partitioner{KaHyPar} yet outperforms it by an average factor of about 9 using 10 threads.
On large instances it typically achieves a self-relative speedup of around 25 using 64 threads.
\Cref{s:conclusion} summarizes the results and outlines directions for future research.

\section{Preliminaries}\label{s:preliminaries}

A \emph{weighted hypergraph} $H=(V,E,c,\omega)$ is defined as a set of vertices $V$ and a set of hyperedges/nets $E$ with vertex weights $c:V \to \mathbb{R}_{>0}$ and net weights $\omega:E \to \mathbb{R}_{>0}$, where each net $e$ is a subset of the vertex set $V$ (i.e., $e \subseteq V$).
The vertices of a net are called its \emph{pins}.
We extend $c$ and $\omega$ to sets in the natural way, i.e., $c(U) :=\sum_{v\in U} c(v)$ and $\omega(F) :=\sum_{e \in F} \omega(e)$.
A vertex $v$ is \emph{incident} to a net $e$ if $v \in e$.
$\mathrm{I}(v)$ denotes the set of all incident nets of $v$.
The \emph{degree} of a vertex $v$ is $d(v) := |\mathrm{I}(v)|$.
The \emph{size} $|e|$ of a net $e$ is the number of its pins.
We call two nets $e_i$ and $e_j$ with different identifiers $i \neq j$ \emph{identical} if they have the same pins.

A \emph{$k$-way partition} of a hypergraph $H$ is a function $\Partition : V \to \{1, \dots, k\}$.
The blocks $V_i := \Partition^{-1}(i)$ of $\Partition$ are the inverse images.
We call $\Partition$ \emph{$\varepsilon$-balanced} if each block $V_i$ satisfies the \emph{balance constraint}: $c(V_i) \leq L_{\max} := (1+\varepsilon) \frac{c(V)}{k}$ for some parameter $\mathrm{\varepsilon} \in (0,1)$.\footnote{With this definition some instances may not admit balanced partitions. This can be remedied by adding $\max_v c(v)$. Here, we use the traditional definition, since it is used by most of the available partitioning software, and this issue does not arise on our benchmark instances.}
A $2$-way partition is also called a \emph{bipartition}.

For each net $e$, $\conset(e) := \{V_i \mid  V_i \cap e \neq \emptyset\}$ denotes the \emph{connectivity set} of $e$.
The \emph{connectivity} $\con(e)$ of a net $e$ is the cardinality of its connectivity set, i.e., $\con(e) := |\conset(e)|$.
A net is called a \emph{cut net} if $\con(e) > 1$, otherwise  (i.e., if $|\mathrm{\lambda}(e)|=1$) it is called an \emph{internal} net.
A vertex $u$ that is incident to at least one cut net is called a \emph{boundary vertex}.
The number of pins of a net $e$ in block $V_i$ is denoted by $\pinsinpart(e,V_i) := |V_i \cap e|$.

Given parameters $\epsilon$, and $k$, and a hypergraph $H$, the \emph{hypergraph partitioning problem} is to find an $\epsilon$-balanced $k$-way partition $\Partition$ that minimizes a certain objective function on the cut nets of $H$.
In this paper we focus on the \emph{connectivity metric} $(\lambda - 1)(\Pi) := \sum_{e \in E} (\lambda(e) - 1) \: \omega(e)$.

Given a $k$-way partition $\Partition$, moving $u$ from its block $\Partition(u)$ to $V_i$ improves the connectivity metric by
$g(u,i) := \omega(\{ e \in I(u) \mid \pinsinpart(e, \Partition(u)) = 1 \}) - \omega(\{e \in I(u) \mid \pinsinpart(e, V_i) = 0 \})$.
This term is called the \emph{gain} of the move.
Only boundary vertices can have positive gains.

\emph{Contracting} the vertex pair $(u,v)$ removes $v$ from all nets $e \in I(u)\cap I(v)$ and replaces $v$ with $u$ in all nets $e \in I(v) \setminus I(u)$.
The weight of $u$ becomes $c(u) := c(u) + c(v)$.
We refer to $u$ as the \emph{representative} and $v$ as the \emph{contracted vertex}.
We also say $v$ is contracted onto $u$, or just $v$ is contracted when the description does not need the representative.
Vertices that are not yet contracted are called \emph{active}.
\emph{Uncontraction} is the reverse operation of contraction.

An undirected graph is called a \emph{forest} if it does not contain a cycle, and a forest is called a \emph{tree} if it is connected.
A rooted forest is a directed graph whose underlying undirected graph is a forest, with a designated set of root vertices such that all edges point towards the root of their tree.
A set $\contractions$ of contractions for a hypergraph $H=(V,E)$ is called \emph{compatible} if the directed graph $\contractionforest = (V, \{(v,u) \mid (u,v) \in \contractions \})$ with edges pointing from contracted vertex to representative, is a rooted forest.
$\contractionforest$ is called the contraction forest.
In this paper, we only construct compatible sets of contractions.
To represent the forest as an array $\rep$, we store the representative of each vertex, i.e., $\rep[v] = u$ for $(u,v) \in \contractions$.
If $v$ was not contracted then $\rep[v] = v$.
These are the roots of $\contractionforest$.
The \emph{ancestors} of $v$ are the vertices on the unique path towards the root of its tree.
The \emph{children} of $v$ are all vertices $w \in V \setminus{v}$ with $\rep[w] = v$, and the \emph{descendants} of $v$ are the vertices in the subtree rooted at $v$.
Vertices $v_1, v_2$ are \emph{siblings} if they are children of the same vertex, i.e., $v_1 \neq \rep[v_1] = \rep[v_2] \neq v_2$.

\section{Related Work}\label{s:relatedWork}

The literature contains a large number of different hypergraph partitioning algorithms, most of which follow the multilevel paradigm
and are either based on recursive bipartitioning (RB) or direct $k$-way partitioning. In recursive bipartitioning, the input hypergraph
is recursively split into two blocks until $k$ blocks are obtained. In direct $k$-way partitioning, the initial partition of the coarsest
hypergraph is already $k$-way. Direct $k$-way achieves better quality~\cite{SimonTeng97, kPaToH, KAHYPAR-K}, but is more difficult to implement because
$k$-way refinement is needed.

The most notable sequential hypergraph partitioners are PaToH (RB)~\cite{PATOH}, which is extremely fast, hMetis
(RB and direct $k$-way)~\cite{HMETIS, HMETIS-K}, the hypergraph version of the well-known Metis graph partitioning algorithms,
as well as KaHyPar (direct $k$-way)~\cite{KAHYPAR-HFC} which aims at very high partition quality.
Parallel algorithms as well as KaHyPar are discussed in more detail later in this section, as these are the most relevant to our contribution.
For a more extensive overview, including non-multilevel algorithms, we refer to existing literature~\cite{ALPERT-SURVEY,GRAPH-SURVEY,PAPA-MARKOV,KAHYPAR-DIS}.

\subparagraph*{Refinement.}

For the refinement phase, most sequential partitioners use the Fiduccia-Mattheyses (FM)~\cite{FM} local search heuristic to improve partitions at each level.
FM repeatedly performs a feasible vertex move with the highest gain (which may be negative), and then returns the best observed solution.
Allowing moves with negative gain enables FM to escape local minima to some extent.
To speed up local search, it is often only initialized with boundary vertices.
FM has been deemed notoriously difficult to parallelize due to the serial move order~\cite{MT-METIS-REFINEMENT}, and is known to be P-complete~\cite{PHARD}.

Label propagation~\cite{HMETIS-K, LABEL_PROPAGATION, PARHIP, PARKWAY-2} is a refinement algorithm that is easier to parallelize.
Vertices are visited in random order (in parallel).
For each vertex, the move with highest positive gain (based on the current view) is performed immediately.
Label propagation lacks the ability to escape local minima since only moves with positive gain are performed.
A synchronous version is amenable for distributed memory, i.e., moves are only communicated and applied after each round~\cite{PARHIP}.
This can incur oscillation, and even decrease partition quality, since moves influence each other's gains.
In shared-memory settings, the balance constraint can be maintained using atomic fetch-and-add instructions.
In distributed memory this poses a substantially more difficult challenge.
One of the possible approaches is discussed in the next paragraph.

\subparagraph*{Parallel Hypergraph Partitioning.}

\Partitioner{Mt-KaHyPar}~\cite{MT-KAHYPAR} is the first shared-memory multilevel hypergraph partitioning algorithm.
It uses direct $k$-way partitioning, and opts for the traditional $\mathcal{O}(\log(n))$ levels by contracting vertex clusterings.
The coarsening is guided by restricting contractions to vertex clusters computed in a preprocessing phase~\cite{KAHYPAR-CA} using the parallel Louvain algorithm~\cite{PARALLEL-LOUVAIN}.
Initial partitioning is done via parallel recursive bipartitioning with work-stealing.
Refinement uses label propagation and a parallel localized FM algorithm that is based on the \emph{localized multi-try} FM of
the graph partitioners \Partitioner{KaHiP}~\cite{KAFFPA} and \Partitioner{Mt-KaHiP}~\cite{MT-KAHIP}.
The threads perform FM local searches that do not overlap on vertices.
Each thread initializes its search with a few different boundary vertices, and gradually expands around it by claiming neighbors of moved vertices.
When a search terminates, it applies its best local prefix to the globally shared partition, and then continues with a different set of seed vertices.
This yields a global move ordering.
Once all initial boundary vertices have been used as seeds, the gains of the moves in the global ordering are recomputed in parallel, and the best prefix is applied to the partition.
While FM finds better partitions than label propagation, it is more difficult to parallelize efficiently.
Therefore, label propagation is employed first to find easy improvements, so that FM can converge faster and only has to find non-trivial improvements.

\Partitioner{Parkway}~\cite{PARKWAY-2} is a distributed multilevel algorithm based on direct $k$-way partitioning, which uses label propagation refinement.
\Partitioner{Zoltan}~\cite{ZOLTAN} is a distributed multilevel algorithm based on recursive bipartitioning.
It uses a 2D distribution, where each processor only has partial information on the vertices and nets allocated to it.
Each processor performs FM on the vertices for which it holds the majority of non-zero matrix entries out of any processor, using and updating only local information.
\Partitioner{SHP}~\cite{SHP} is a distributed non-multilevel (flat) algorithm which uses a modified objective function to reduce the effect of zero gain moves.
Similar to synchronous label propagation, it first computes the gains for all vertices and target blocks, and then selects the highest gain target block for each vertex.
For each block pair, vertices are paired off for swaps by highest gain first.
This is approximated in a distributed setting by bins for each move direction, with exponentially increasing gain ranges.
Two bins with the same range swap the maximal possible number of vertices (size of smaller bin).
Vertices for which the bin of the other move direction is empty, are moved with a certain probability that attempts to maintain the balance constraint.
\Partitioner{BiPart}~\cite{BIPART} is a deterministic shared-memory multilevel algorithm based on recursive bipartitioning.
For $2$-way refinement, it uses the same algorithm as \Partitioner{SHP} but simply uses a vector sorted by gains for each move direction,
and swaps a prefix of the vectors.

\subparagraph*{KaHyPar.}
We briefly describe the coarsening and uncoarsening of the sequential $n$-level algorithm \Partitioner{KaHyPar}~\cite{KaHyPar-R, KAHYPAR-K} since our parallel algorithms are based on them.

The coarsening algorithm proceeds in \emph{passes} until the hypergraph only has $160 \cdot k$ vertices remaining, or no vertex was contracted in the last pass.
In each coarsening pass, the remaining active vertices are visited in random order.
For each vertex $u$, three steps are performed.
First, the highest rated neighbor $v$ according to the \emph{heavy-edge} rating function~\cite{PATOH,KAHYPAR-CA,HMETIS}
\[r(u, v) := \sum_{e \in I(u) \cap I(v)} \frac{\omega{(e)}}{|e| - 1},\]
is selected to be contracted onto $u$.
This function prefers vertices that share a large number of heavy nets of small size with $u$.
Creating too heavy vertices impairs initial partitioning algorithms in finding balanced partitions, and may even make it impossible.
Hence, the selection is restricted to candidates $v$ with $c(u) + c(v) \leq c_{\max} := \frac{c(V)}{160 \cdot k}$.
Secondly, the contraction $(u,v)$ is performed immediately, so that later iterations already see the aggregated vertex $u$.
Third, nets of $u$ that become identical through the contraction (have the same set of pins) are removed, except for one representative at which the weights of the other nets are aggregated.
Nets consisting of a single pin are discarded.

After computing an initial $k$-way partition, the contractions are undone in the reverse order in which they were performed.
A restored vertex is assigned to the same block in the partition as its representative.
After each uncontraction, localized FM search is performed around the two vertices of the reverted contraction.

\section{Parallel $n$-level Hypergraph Partitioning}\label{s:main}
\begin{algorithm2e}
	\KwIn{Hypergraph $H = (V,E)$, number of blocks $k$}
	\KwOut{$k$-way partition $\Partition$ of $H$}
	\SetEndCharOfAlgoLine{}
	\caption{Parallel $n$-level Hypergraph Partitioning}\label{pseudocode:parallelization}

	$\forall v \in V: \rep[v] \gets v$ \tcp*[r]{\small initialize empty contraction forest $\contractionforest$}
	\While() {$|V| > 160 \cdot k$}{
		\ParallelFor(\tcp*[f]{\small coarsening pass}\label{pseudocode:coarsening_pass}) {$u \in V$ in random order}{
			$v \gets \operatorname{arg\,max}_{v} \sum_{e \in I(u) \cap I(v)} \frac{\omega{(e)}}{|e| - 1}$  \tcp*[r]{\small$O(\sum_{e \in I(u)}|e|)$} \label{pseudocode:rating_function}
			\If(\tcp*[f]{\small Section~\ref{sec:parallel_contractions}} \label{pseudocode:safe_contractions}) {$(u,v)$ can be safely added to $\contractionforest$} { \label{pseudocode:compatible_contraction}
				$\rep[v] \gets u$ and contract $v$ onto $u$ \tcp*[r]{\small Section~\ref{sec:dynamic_hypergraph}}
			}
		}
	}

	$\Partition \gets$ \FuncSty{initialPartition}($H, k$) \tcp*[r]{\small Section~\ref{sec:initial_partitioning}}
	$\batches = \langle B_1, \dots, B_l \rangle \gets \FuncSty {constructBatches}(\contractionforest)$ \tcp*[r]{\small Section~\ref{sec:uncoarsening}}\label{pseudocode:start_uncoarsening}
	\For(\tcp*[f]{\small $|B| \approx \maxbatchsize$}) {$B \in \batches$}{
		\ParallelFor() {$v \in B$}{
			uncontract $v$ from $\rep[v]$ \tcp*[r]{\small Section~\ref{sec:dynamic_hypergraph}}
			$\Partition[v] \gets \Partition[\rep[v]]$\;
		}
		improve $\Partition$ with localized refinement on $B$ \tcp*[r]{\small Section~\ref{sec:uncoarsening}}
	}
\end{algorithm2e}

Algorithm~\ref{pseudocode:parallelization} shows the high-level pseudocode of our parallel $n$-level framework.
KaHyPar's approach of contracting and uncontracting vertices in a strict order is inherently sequential which is why we have to relax the $n$-level paradigm.
For the coarsening phase, we parallelize the loop over the active vertices in Line~\ref{pseudocode:coarsening_pass}.
Contractions are performed on-the-fly as in the sequential algorithm.
The challenge addressed in Section~\ref{sec:parallel_contractions} is to keep the contractions compatible (see Line~\ref{pseudocode:safe_contractions}) as well as to determine a schedule for the contraction operations.
The uncoarsening phase starts in Line~\ref{pseudocode:start_uncoarsening}.
We construct a sequence of batches $\batches = \langle B_1, \dots, B_l \rangle$ of contracted vertices, such that $|B_i| \approx \maxbatchsize$, where $\maxbatchsize$ is an input parameter.
Batches are processed one after another, enabling the uncontraction of vertices in subsequent batches.
Vertices in the same batch are uncontracted in parallel.
Note that setting $\maxbatchsize = 1$ corresponds to sequential \Partitioner{KaHyPar}.
The challenge is to identify which vertices can or even must appear in the same batch, which we address in Section~\ref{sec:uncoarsening}.
After each batch, we apply highly-localized refinement algorithms around the vertices in the batch.
The localized refinement algorithms described in Section~\ref{sec:uncoarsening} are adaptions from the algorithms presented in our previous paper~\cite{MT-KAHYPAR}.
In Section~\ref{sec:initial_partitioning}, we additionally introduce an improvement to initial partitioning.
Furthermore, we propose a new low-memory hypergraph data structure in Section~\ref{sec:dynamic_hypergraph}, and describe how to implement contractions and uncontractions on it.
This will reveal certain conditions and intricacies regarding parallelization that must be addressed by the coarsening and uncoarsening algorithms in Sections~\ref{sec:parallel_contractions} and~\ref{sec:uncoarsening}.

\subsection{Dynamic Hypergraph Data Structure}\label{sec:dynamic_hypergraph}

\begin{figure*}
	\centering
	\includegraphics[width=\textwidth]{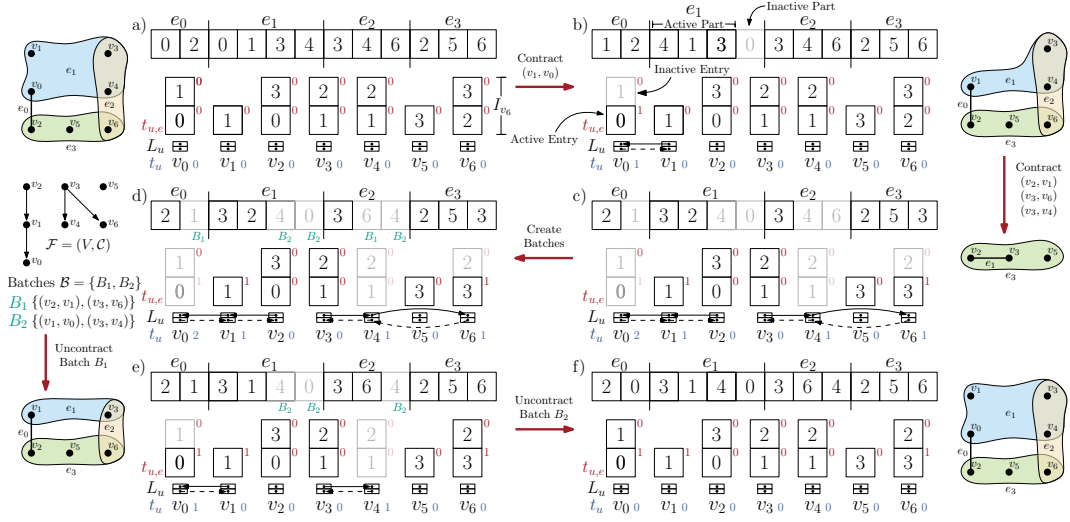}
	\caption{Illustrates the contraction and batch uncontraction operations applied to the dynamic hypergraph data structure.}\label{fig:nlevel_example}
\end{figure*}

To support concurrent on-the-fly contraction and uncontraction, we use a dynamic hypergraph data structure that improves the approach in KaHyPar~\cite[p.~100]{KAHYPAR-DIS}.
In KaHyPar, the pins of the nets are represented as an adjacency array (the sub-range storing the pins of a certain net is called its pin-list), whereas the incident nets of the vertices are represented using adjacency-lists, i.e., as a separate vector for each vertex.
Because of the adjacency-lists, a contraction $(u,v)$ entails copying $I(v) \setminus I(u)$ to $I(u)$.
In the worst case, this can lead to quadratic memory usage and is therefore not practical for large or skewed hypergraphs.
Instead, we propose a new data structure for storing incident nets that enables concurrent (un)contractions without allocating additional memory.
In preliminary experiments, this data structure was roughly $5\%$ slower than KaHyPar's approach due to iteration overheads, but allows us to handle larger and skewed instances.

The key idea is to remove $I(u) \cap I(v)$ from $I(v)$ (instead of adding $I(v) \setminus I(u)$ to $I(u)$).
$I(u)$ is obtained by iterating over both the representation of the current $I(u)$ and the remaining entries of $I(v)$.
For each vertex $u \in V$, we store an array $I_u$ which is initialized with the incident nets of $u$ on the input hypergraph.
We organize all vertices contracted onto $u$ as well as $u$ itself in a doubly-linked list $L_u$, so that the \emph{current state} for $I(u)$ is obtained by iterating over all $I_w$ arrays for $w \in L_u$.
When contracting $v$ onto $u$, we remove any incident net of $u$ from the arrays $I_w$ for $w \in L_v$ and append $L_v$ to $L_u$.
For storing the pin-lists of nets, we use an adjacency array as in KaHyPar.
The data structure and all operations, which we describe in more detail in the following, are illustrated with an example of multiple contraction and uncontraction steps in Figure~\ref{fig:nlevel_example}.
In each step, the top part shows the current state of the pin-lists, the bottom part shows the incident net arrays $I_w$ and lists $L_w$.

\subparagraph*{Remove and Restore Incident Nets.}

To remove and later restore entries from an incident net array $I_w$, we additionally store a counter $t_w$ which counts in
how many contractions $I_w$ was modified, as well as a marker $t_{w,e}$ for each entry of $I_w$.
The counter and markers are initially set to zero.
Entries with markers $\geq t_w$ are \emph{active}, i.e., were not removed yet.
To remove a set $X$ of entries from $I_w$, we increment $t_w$ and iterate over the previously active entries of $I_w$ (now marked with $t_w -1$).
If the entry is not in $X$, we set its marker to $t_w$.
Otherwise, we swap the entry and its marker to the end of the active part but keep its marker at $t_w - 1$, thereby marking the
entry inactive.
This maintains the invariant that the entries of $I_w$ are sorted by decreasing markers, so that iterating over active entries of $I_w$ has no overhead.
In particular the iterator for $I(u)$ has a complexity of $\Oh{|I(u)| + |\doublelinkedlist{u}|}$.

To restore entries, we just decrement $t_w$, so that we consider entries marked with $t_{w,e} = t_w - 1$ as active again.
The restore operations must be performed in reverse order of the remove operations, in order to restore the correct entries.
This is guaranteed in our scenario, since a vertex can only be uncontracted once its ancestors in
$\contractionforest$ have been uncontracted.

In Figure~\ref{fig:nlevel_example} b we contract $v_0$ onto $v_1$.
We remove $e_1$ from the incident net array of $v_0$, but keep $e_0$.
Therefore, we increment $t_0$ to $1$, and set $t_{0, 0}$ to $1$, while leaving $t_{0, 1}$ at $0$.
In step f (which uncontracts $v_1$) we decrement $t_{0}$ to $0$ and thus mark $e_1$ as active again in the incident net array of $v_0$.

\subparagraph*{Contraction Operation.}

\begin{algorithm2e}
	\KwIn{Contraction $(u,v)$}
	\SetEndCharOfAlgoLine{}
	\caption{Contraction Operation}\label{pseudocode:contraction}

	$c(u) \gets c(u) + c(v)$\;
	\For(\tcp*[f]{\small edit pin-lists}){$e \in I(v)$}{
		\If(\tcp*[f]{\small $e \in I(u) \cap I(v)$} \label{pseudocode:search_u}) { $u$ is contained in the pin-list of $e$ } {
			remove $v$ from the pin-list of $e$\;\label{pseudocode:contraction_remove}
			mark $e$ in bitset $X$\;
		} \Else(\tcp*[f]{\small $e \in I(v) \setminus I(u)$}) {
			replace $v$ by $u$ in the pin-list of $e$\;\label{pseudocode:contraction_replace}
		}
	}

	\For(\tcp*[f]{\small edit incident net arrays}) {$w \in L_v$} {
		remove all active entries marked in $X$ from $I_w$\;\label{pseudocode:edit_incident_net_array}
	}
	append $L_v$ to $L_u$\;\label{pseudocode:append_incident_net_array}
\end{algorithm2e}

Algorithm~\ref{pseudocode:contraction} shows the pseudocode for contracting a vertex $v$ onto another vertex $u$.
To edit the pin-lists, we iterate over the incident nets $e \in I(v)$ of $v$ and search for the position of $u$ in $e$ (see Line~\ref{pseudocode:search_u}).
If we do not find $u$, we replace $v$ by $u$ in the pin-list of $e$ in Line~\ref{pseudocode:contraction_replace}
(e.g., $v_0$ is replaced by $v_1$ in $e_0$ in Figure~\ref{fig:nlevel_example} b).
Otherwise, we swap $v$ to end of the active part of $e$ and decrement the current size of $e$ (see Line~\ref{pseudocode:contraction_remove}
and pin $v_0$ of $e_1$ in Figure~\ref{fig:nlevel_example} b), as well as mark $e$ in a bitset $X$.
We use this bitset to remove all nets $e \in I(u) \cap I(v)$ from the incident net arrays $I_w$ for all $w \in L_v$ in
Line~\ref{pseudocode:edit_incident_net_array}.

To enable concurrent contractions, we use a separate lock for each net to synchronize edits to the pin-lists.
In Line~\ref{pseudocode:append_incident_net_array}, the set $I(u)$ may change due to concurrent contractions onto $u$, which is why it is not thread-safe to initialize $X$ by iterating over $I(u)$.
If multiple vertices that are contracted concurrently onto $u$ share a net $e$, only the first pin-list edit of $e$ can do the replacement (if $u$ was not already in $e$).
All subsequent edits of $e$ correctly remove their pins and mark $e$ in their thread-local bitset $X$.

Operations on the incident net arrays $I_w$ for $w \in L_v$ are not synchronized (see Line~\ref{pseudocode:edit_incident_net_array}),
since only contractions of descendants of $v$ in $\contractionforest$ can modify these arrays.
These must be finished before the contraction of $v$ starts, and our algorithm in Section~\ref{sec:parallel_contractions} guarantees this.
Additionally, we use a separate lock for each vertex $u \in V$ to synchronize modifications to $L_u$ and $c(u)$.
If $c(u) + c(v)$ exceeds the maximum vertex weight $c_{\max}$, we discard the contraction.

\subparagraph*{Uncontraction Operation.}

\begin{algorithm2e}
	\KwIn{Contraction $(u,v)$}
	\SetEndCharOfAlgoLine{}
	\caption{Uncontraction Operation}\label{pseudocode:uncontraction}

	$\Partition(v) \gets \Partition(u)$\;
	restore the sublist $L_v$ from $L_u$\;\label{pseudocode:splice}
	\For {$w \in L_v$} {
		$t_w \gets t_w - 1$\; \label{pseudocode:restore_incident_nets}
		\ForAll(\tcp*[f]{\small $t_{w,e} \ge t_w$}) { active entries $e \in I_w$ } {
			\If(\tcp*[f]{\small $e \in I(u) \cap I(v)$}\label{pseudocode:uncontraction:case_distinction}) { $t_{w,e} = t_w$ } {
				restore $v$ in the pin-list of $e$\; \label{pseudocode:uncontraction_restore}
			} \Else(\tcp*[f]{\small $e \in I(v) \setminus I(u)$}) {
				replace $u$ by $v$ in the pin-list of $e$\; \label{pseudocode:uncontraction_replace}
			}
		}
	}
	$c(u) \gets c(u) - c(v)$\;

\end{algorithm2e}

Algorithm~\ref{pseudocode:uncontraction} shows the pseudocode for uncontracting a vertex
$v$ that is contracted onto a vertex $u$.
To restore $L_v$ from $L_u$ in Line~\ref{pseudocode:splice}, we additionally store the last vertex in $L_v$ at the time $v$ is contracted.
To restore the incident nets of $v$ that were removed, we iterate over all vertices $w \in L_v$ and decrement the counter $t_w$ in Line~\ref{pseudocode:restore_incident_nets}.
This reactivates all entries of $I_w$ that became inactive due to contracting $v$, i.e., had marker $t_{w,e} = t_w$.
The other active nets are marked with $t_{w,e} > t_w$, which were not incident to $u$ at the time of the contraction and thus not removed.
To restore the pin-lists, we iterate over all active nets $e \in I_w$ and if $t_{w,e} = t_w$, we restore $v$ from the inactive
part of the pin-list of $e$ in Line~\ref{pseudocode:uncontraction_restore} (see $v_4$ in $e_2$ in
Figure~\ref{fig:nlevel_example} f). Otherwise, if $t_{w,e} > t_w$,
we replace $u$ by $v$ in the pin-list of $e$ in Line~\ref{pseudocode:uncontraction_replace}
(see $v_0$ replacing $v_1$ in $e_0$ in Figure~\ref{fig:nlevel_example} f).

In the sequential setting of KaHyPar~\cite{KAHYPAR-DIS} contractions are undone in the reverse order in which they were performed, so if $v$ was removed, it is the first entry in the inactive part of $e$.
In this case it suffices to increment the current size of $e$ to restore $v$ in Line~\ref{pseudocode:uncontraction_replace}.
In the parallel setting, we perform all uncontractions in the current batch $B$ in parallel, so $v$ can be anywhere in the inactive part of $e$'s pin-list.
After constructing the batches, we sort each pin-list (in particular the inactive entries) by the batches in which the pins are uncontracted, see net $e_2$ in Figure~\ref{fig:nlevel_example} d.
Then all pins of $e$ that have to be restored in the current batch can be activated simultaneously by appropriately raising the current size of $e$, as seen in parts e and f of Figure~\ref{fig:nlevel_example}.
Only one thread that triggers the restore case on a net performs the restore operation.
We ensure this with an atomic test-and-set instruction on a bit for this net, which we reset after each batch.

\subparagraph*{Further Implementation Details.}

While the description references a separate incident net array for each vertex, they are actually organized as an adjacency array as well.
To speed up iteration over incident nets, we store a second doubly-linked list where vertices $w$ without active entries in $I_w$ are removed.
This improves the iterator's complexity to $\Oh{|I(u)|}$.
Since the edit operations that require locks are very fine-grained, we implement locking with spinlocks using atomic test-and-set operations.
This applies to all locks used in this paper.

\subsection{Parallel $n$-level Coarsening}\label{sec:parallel_contractions}
To outline the parallelization idea, we assume the contraction forest $\contractionforest$ is known in advance, before we describe how to lift this restriction.
A vertex can be contracted as soon as all of its children in $\contractionforest$ have been contracted.
To obtain parallelism, such vertices can be contracted independently, i.e., we traverse $\contractionforest$ in bottom-up fashion.
The contracted hypergraph will be the same regardless of exact execution order, only the data structure representation may differ, i.e., which pin was replaced by its representative and which pin was removed.

In the following, we describe how we dynamically extend $\contractionforest$ in a thread-safe manner that still
enables this parallelization scheme, while simultaneously performing the contraction operations defined by $\contractionforest$.
For finding good contractions, we use the same algorithm as sequential KaHyPar but parallelize the iteration
over active vertices (c.f. Line~\ref{pseudocode:coarsening_pass} in Algorithm~\ref{pseudocode:parallelization}) and run the algorithm described below instead of performing the contraction right away (as shown in Line~\ref{pseudocode:compatible_contraction}).
While the rating function in Line~\ref{pseudocode:rating_function} uses potentially outdated information due to concurrent contractions, this has negligible impact on solution quality.
The same observation has been made for similar clustering and community detection algorithms~\cite{PARALLEL-LOUVAIN, SLM, MT-KAHYPAR}.

\subparagraph*{Handling Contraction Dependencies.}

Performing a contraction $(u,v)$ does not break compatibility with existing contractions, if it satisfies the following two conditions:

\begin{enumerate}
	\item $\contractionforest$ must remain a rooted forest
	\item the contraction of $u$ must not be started yet.
\end{enumerate}

More precisely, adding edge $(v,u)$ to $\contractionforest$ must not induce a cycle and $v$ must still be a root, i.e., $\rep[v] = v$.
If $u$ is already contracted or processed by a different thread, contracting $v$ onto $u$ would introduce inconsistencies as $u$ may already have been replaced in some of its incident nets by $\rep[u]$.
This has two consequences:
First, if the contraction of $u$ has started, we instead contract $v$ onto a suitable ancestor of $u$.
Secondly, if there are unfinished contractions onto $v$, we cannot contract $v$ right away.
Note that we explicitly allow multiple concurrent contractions onto the same vertex.

We use one lock per vertex, as well as a zero-initialized array $\pending$, where $\pending[x]$ stores the number of vertices $y$ with $\rep[y] = x$ whose contraction is not finished.
If $\pending[x] = 0$, it is safe to contract $x$.
If additionally $\rep[x] \neq x$, we assume that the contraction of $x$ onto $\rep[x]$ has started.
The entries $\rep[x]$ and $\pending[x]$ are only modified while holding the lock for $x$.
The following algorithm takes a desired contraction $(u,v)$ as input, and ensures that at some point a contraction $(u', v)$
is applied to the dynamic hypergraph data structure, where $u'$ is either $u$ or an ancestor of $u$. If some other thread already set $\rep[v]$, we can safely discard $(u,v)$.
Note that in most cases $u'$ is just $u$, i.e., we perform the requested contraction.

First, we acquire the lock for $v$, so that no other thread can write to $\rep[v]$.
If $\rep[v] \neq v$, we discard the contraction $(u,v)$, as another thread has already selected a representative for $v$ and will run this algorithm.
Otherwise, we walk the path towards the root of $u$'s tree in $\contractionforest$ by chasing the $\rep$ entries to find the lowest ancestor $u'$ of $u$, for which either $\rep[u'] = u'$ or $\pending[u'] > 0$.
This guarantees that the contraction of $u'$ has not started.
If we find $v$ on this path, we discard the contraction $(u,v)$, as it would add a cycle to $\contractionforest$.

If we did not find a cycle, we acquire the lock for $u'$ and check $\rep[u']$ and $\pending[u']$ again.
If they changed, we release $u'$ and keep walking up to find an ancestor of $u'$.
Otherwise, $u'$ is the desired candidate.
We still finish the walk up to the root to check for cycles.
If no cycles are found, we set $\rep[v] \gets u'$ and $\pending[u'] \gets \pending[u'] + 1$, and release the locks for $v$ and $u'$.
To avoid deadlocks, we acquire the lock of the vertex with lower id first.
If $u' < v$, we have to intermediately release the lock for $v$, so we re-check whether $\rep[v] = v$ and discard the contraction if this is not the case.
If the contraction is discarded at any stage, we release all appropriate locks.

Now, we try to start the contraction operation for $(u', v)$.
First, we acquire the lock for $v$ again.
If $\pending[v] > 0$, we are finished, as we transfer the responsibility for contracting $(u',v)$ to the thread that reduces $\pending[v]$ to $0$.
If $\pending[v] = 0$, we release the lock and start the contraction operation on $(u',v)$.
After the contraction, we decrement $\pending[u']$ by $1$ -- with acquired lock for $u'$.
If this reduces $\pending[u']$ to $0$ and $\rep[u'] \neq u'$, this thread is responsible for performing the contraction $(\rep[u'], u')$, so we recursively apply this process.

Algorithm~\ref{algo:contract_deps} in Appendix~\ref{appendix:contract_deps} shows the pseudocode for the algorithm described above.

\subparagraph*{Removing Identical Nets.}

There is one last detail left for the coarsening phase.
We remove identical and single-vertex nets after each coarsening pass, using our parallel algorithm from~\cite{MT-KAHYPAR} adapted to the dynamic data structure.
Removing this redundant information speeds up the other algorithmic components of the framework.
Doing this on-the-fly as in sequential KaHyPar would introduce additional dependencies for the batches in the uncoarsening phase, which is why we decided against it.
Efficient, parallel, on-the-fly identical net removal might be an avenue for future research.

\subsection{Initial Partitioning}\label{sec:initial_partitioning}

Initial partitioning is done via recursive multilevel bipartitioning with work-stealing as described in~\cite{MT-KAHYPAR} but using parallel $n$-level coarsening and uncoarsening as described in Section~\ref{sec:parallel_contractions} and~\ref{sec:uncoarsening}.
To obtain a $2$-way partition of the coarsest hypergraph, we choose the best bipartition computed by a pool of $9$ flat bipartitioning algorithms.
Instead of running them $20$ times each as in~\cite{MT-KAHYPAR}, we perform at least $5$ runs and at most $20$.
After $5$ runs, we only run an algorithm again, if it is likely to improve the best solution $\Partition^*$ found so far.
We estimate this based on the arithmetic mean $\mu$ and standard deviation $\sigma$ of the connectivity values achieved by that algorithm so far, using the $95\%$ rule.
Assuming the connectivity values follow a normal distribution, roughly $95\%$ of the runs will fall between $\mu - 2 \sigma$ and $\mu + 2 \sigma$.
If $\mu - 2 \sigma > (\lambda -1)(\Partition^*)$, we do not run this flat algorithm again.

\subsection{Parallel $n$-level Uncoarsening and Refinement}\label{sec:uncoarsening}

For the uncoarsening phase, our goal is to create a sequence of batches $\batches = \langle B_1, \ldots, B_l \rangle$, where \batches~is a partition of the contracted vertices into disjoint sets such that $\forall B \in \batches: |B| \approx \maxbatchsize$.
$\maxbatchsize$ is an input parameter that interpolates between scalability (high values) and solution quality ($\maxbatchsize = 1$ corresponds to KaHyPar's $n$-level approach).
Each batch $B_i$ will be chosen such that we can uncontract the vertices $v \in B_i$ in parallel.
Refinement is applied after each batch.
Processing $B_i$ will resolve the last dependencies required to uncontract the next batch $B_{i + 1}$.
Clearly, the uncontraction of a vertex $v$ can only start once the uncontraction of $\rep[v]$ is finished, i.e., all of its ancestors are uncontracted.
Therefore, we construct the batches via a top-down traversal of $\contractionforest$.
However, due to the way we perform the replacement edits of the pin-lists, we have introduced additional ordering dependencies between siblings in $\contractionforest$.

\subparagraph*{Handling Sibling Uncontraction Dependencies.}

\begin{figure}
  \centering
	\includegraphics[width=0.75\textwidth]{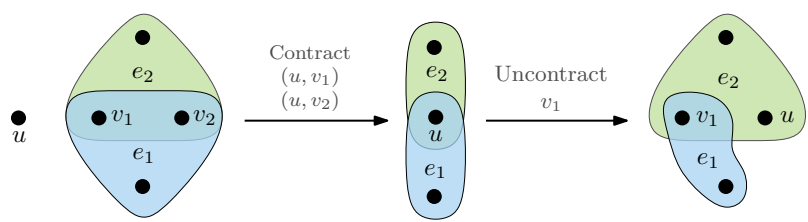}
	\caption{\normalfont Inconsistent state due to concurrent contraction of both $v_1$ and $v_2$ onto $u$, where $u$ replaces $v_1$ in $e_1$ and $v_2$ in $e_2$. By uncontracting $v_1$ before $v_2$, it replaces $u$ in $e_1$ again, but $u$ should still be incident to $e_1$, since $v_2$ is still contracted onto $u$.}\label{fig:uncontraction_order_entanglement}
\end{figure}

Consider the scenario in Figure~\ref{fig:uncontraction_order_entanglement}, with vertices $u, v_1, v_2 \in V$ with $\rep[v_1] = \rep[v_2] = u$, and two nets $e_1, e_2 \in E$
with $v_1, v_2 \in e_i$ but $u \notin e_i$.
If the contractions of $v_1$ and $v_2$ happen at the same time, and $u$ replaces $v_1$ in $e_1$ and $v_2$ in $e_2$,
then $v_2$ gets removed in $e_1$, and $v_1$ gets removed in $e_2$ (moved to the inactive part).
If we uncontract $v_1$ in an earlier batch than $v_2$, then $u$ would be incident to $e_2$ but not $e_1$ until $v_2$ is uncontracted, since we replace $u$ by $v_1$ in $e_1$ but not yet by $v_2$ in $e_2$.
This is an inconsistent state because $v_2$ is in $e_1$ and was contracted onto $u$, and thus $u$ should be incident to $e_1$.
In this case refinement algorithms would yield incorrect results.
Hence, we enforce that all sibling contractions that happened \emph{at the same time} must be reverted in the same batch.
A similar argument holds for the case of $v_1$ being contracted strictly earlier than $v_2$ (no time overlap).
Here, $v_2$ must be uncontracted in an earlier or the same batch as $v_1$.

To detect time overlaps, we atomically increment a counter before starting and after finishing a contraction operation.
For each contracted vertex $v$, this yields an interval $[s_v, e_v]$ with start time $s_v$ and end time $e_v$.
If the intervals of two vertices overlap, we assume they were contracted at the same time,
otherwise one is strictly earlier than the other.
Among siblings, we need to compute the transitive closure of vertices with overlapping intervals, as well as order them decreasingly if one is strictly earlier than the other.
Since comparing for equality with interval overlaps is not transitive, we instead sort them in decreasing order of $e_v$.
Then, the vertices in a transitive closure are ordered consecutively and can be found with a rightward sweep from the first interval, by checking whether the next interval overlaps with the union of intervals in the closure so far and extending the union interval if they overlap.

\subparagraph*{Batch Construction Algorithm.}

We traverse $\contractionforest$ top-down in Breadth-First-Search (BFS) order, using two FIFO queues $Q$ for the current BFS layer and $Q'$ for the next layer.
Additionally, we have a batch $\currentbatch$ that we are currently adding contracted vertices to.
If $|\currentbatch| \geq \maxbatchsize$ or we get to the next level in the BFS, we append $\currentbatch$ to $\batches$, and proceed with a new empty batch.

$Q$ and $Q'$ store elements $(u, v_i)$, where $u$ is a vertex and $v_i$ is the $i$-th child of $u$ in the sorted order described above.
For elements in $Q$, we maintain the invariant that $u$ is uncontracted in a batch before $\currentbatch$, so that its children can be added to $\currentbatch$.
Furthermore, $v_i$ is the first child of $u$ that has not been added to any batch yet.
To initialize $Q$, we insert all entries $(r, w_1)$, where $r$ is a root of $\contractionforest$ and $w_1$ is the first child of $r$.

We pop elements from $Q$ until it is empty, and then swap it with $Q'$.
Now, let $(u,v_i)$ be the current element we popped from $Q$, and let $T$ denote the transitive closure of $v_i$,
as described in the previous paragraph.
For each $v \in T$, we add $v$ to $\currentbatch$, and push $(v, w_1)$ to $Q'$, where $w_1$ is the first child of $v$.
Additionally, we push $(u, v_j)$ to the end of $Q$, where $v_j$ is the first child of $u$ outside $T$, if any.
The reason for this reinsertion is to minimize the number of contractions in the same batch which have the same representative.
This reduces synchronization overheads during uncontraction operations and reduces the overlap of local searches.
The complexity of this algorithm is $\Oh{|V|}$.

We obtain parallelism by traversing the different trees of $\contractionforest$ concurrently.
$\contractionforest$ has as many trees as vertices left in the coarsest hypergraph, so at least around $160 \cdot k$.
To approximate a BFS order across trees, we perform one BFS per thread, which is initialized with the different roots (and first children) assigned to the thread.
The threads collaborate on filling batches, and we keep multiple batches open, as threads may progress at different rates.
The parallelization is work-efficient in the theoretical sense, as the work is still $\Oh{|V|}$.
The span of the algorithm is linear in the maximum tree size of $\contractionforest$.

\subparagraph*{Refinement.}

We perform refinement after uncontracting a batch of vertices.
To make this efficient, the refinement should be \emph{localized}, i.e., focus on areas close to the uncontracted vertices.
For label propagation, we achieve this by only iterating over vertices in the batch.\looseness=-1

The parallel FM algorithm presented in our previous work~\cite{MT-KAHYPAR} is already localized, i.e., each thread expands its search around a few seed vertices.
Instead of considering all boundary vertices as seeds, we only use vertices from the current batch as seeds.
Note that the searches may expand to vertices that are not in the batch.
Furthermore, we do not use the proposed parallel gain recalculation technique, since it has a substantial startup overhead and we expect only few vertices to be moved to a different block. A simple sequential gain recalculation by re-applying the moves one by one suffices in this case.

We complement the localized refinement with a refinement pass on the entire hypergraph after contractions from a full coarsening pass (one iteration over vertices) are reverted.
This corresponds to the refinement available in traditional multilevel algorithms.

To improve scalability of the direct $k$-way refinement, we actually uncontract multiple batches until the number of boundary vertices in the uncontractions exceeds a threshold $\beta = \max(\maxbatchsize, 50 \cdot t)$, where $t$ is the number of threads, and only then perform localized refinement.
For initial partitioning, we set $\beta = 0$, since the nested parallelism available from recursive bipartitioning suffices.

\subparagraph*{Gain Table Maintenance.}

The localized $k$-way FM refinement algorithm moves vertices greedily according to a gain value.
The gain $g(u,i)$ is the decrease in the connectivity metric, if vertex $u$ were moved to block $i$.
Moving a vertex affects the gains of its neighbors.
Recalculating the gain of each neighbor from scratch after a move incurs too much overhead.
Therefore the gain values are maintained in a table, and updated throughout the algorithm~\cite{MT-KAHYPAR}.

Recall that $\pinsinpart(e,i)$ is the number of pins of net $e$ in block $V_i$, the connectivity $\lambda(e)$ of a net $e$ is $\lambda(e) = |\{V_i \mid \pinsinpart(e,i) > 0\}|$, and the objective function is $\sum_{e \in E} \omega(e) (\lambda(e) - 1)$.
The following two terms are stored separately to calculate the gain.

\begin{equation}
\begin{aligned}
b(u) &\mathrel{:}= \omega(\{e \in \incnets(u) \mid \pinsinpart(e, \Partition(u)) = 1  \}) \\
p(u, i) &\mathrel{:}= \omega(\{e \in \incnets(u) \mid \pinsinpart(e, i) \geq 1\})
\end{aligned}
\end{equation}

The gain $g(u,i)$ of moving a vertex $u$ to block $i$ is then computed as follows.

\begin{equation}
\begin{split}
g(u, i) &= b(u) - \omega(\{e \in \incnets(u) \mid \pinsinpart(e, i) = 0\}) \\
		&= b(u) - \omega(\incnets(u)) + p(u,i)
\end{split}
\end{equation}

The previous algorithm initialized the gains on each level from scratch~\cite{MT-KAHYPAR}.
Since this would incur too much overhead with $O(|V|)$ levels, we instead update the gain entries based on the uncontraction operations described in Section~\ref{sec:dynamic_hypergraph}.

Consider a contraction $(u,v)$ that should be reversed.
We update the above terms for $u$ and $v$ based on updates to the $\pinsinpart(e, i)$ values of nets $e \in \incnets(v)$.
While updates to $v$ are done exclusively by one thread, updates to $u$ can happen from multiple threads, which is why we use atomic fetch-and-add instructions in this case.
The values $b(v)$ and $p(v,i)$ are initialized to zero.
The corresponding values for $u$ are still correct from the previous levels.
Analogously to the uncontraction operation, on each incident net $e$ of $v$, we distinguish two cases (c.f. Line~\ref{pseudocode:uncontraction:case_distinction} of Algorithm~\ref{pseudocode:uncontraction}) for the gain updates due to uncontracting $v$:
(i) whether $v$ was replaced by $u$ in $e$, or (ii) whether $u$ and $v$ were both incident to $e$ before the contraction.

If $v$ was replaced by $u$, we subtract $\omega(e)$ from $p(u,i)$ and add it to $p(v,i)$ for all $i \in \Lambda(e)$.
Recall that $\Lambda(e) = \{ i \mid \pinsinpart(e,i) \geq 1\}$.
Additionally, if $\pinsinpart(e, \Partition(u)) = 1$, we subtract $\omega(e)$ from $b(u)$ and add it to $b(v)$.
In this case, $\pinsinpart(e, \Partition(u))$ does not change, since $v$ is now a pin of $e$, but $u$ no longer is.

In the second case, where $u$ and $v$ are both incident to $e$ after the uncontraction, we increase $\pinsinpart(e, \Partition(u))$ by one since $v$ is assigned to $\Partition(u)$ and now both $u$ and $v$ are pins of $e$ in $\Partition(u)$.
If this increased the value to $2$, we have to reduce $b(u)$ by $\omega(e)$, since moving $u$ out of $\Partition(u)$ would no longer remove $\Partition(u)$ from $\Lambda(e)$.
Additionally, we add $\omega(e)$ to each $p(v,i)$ for $i \in \Lambda(e)$.

$\Lambda(e)$ does not change during uncontractions, since $\pinsinpart(e, i)$ is only modified in the second case, where $\pinsinpart(e,i) \geq 1$.
Therefore, iteration over $\Lambda(e)$ is thread-safe.
Modification and reads on $\pinsinpart(e, \Partition(u))$ are thread-safe because they are protected by a net-specific lock for $e$.

The parallel setting introduces one more intricacy.
In the second case, $u$ might have been replaced in the active part of $e$ by some $v' \neq v$ due to a concurrent uncontraction $(u, v')$.
Therefore, we search for a pin $v'$ in the active part of $e$ with $\Partition(v') = \Partition(u)$ and $v' \neq v$, and then update $b(v')$ instead of $b(u)$.
If $u$ was not replaced, we simply find $v' = u$.

Since we do this while holding the lock for $e$, we can guarantee that if $\pinsinpart(e, \Partition(u))$ is incremented to $2$, there are only two vertices of $\Partition(u)$ in the active part of $e$.

%
%
%

\section{Experiments}\label{s:experiments}

\begin{filecontents*}{experiments/scalability_experiments/speedup.dat}
speedUpTotalTimeP4S0 = 3.7
speedUpTotalTimeP4S100 = 3.7
speedUpTotalTimeP16S0 = 11.7
speedUpTotalTimeP16S100 = 12.3
speedUpTotalTimeP64S0 = 22.6
speedUpTotalTimeP64S100 = 25
speedUpCoarseningP4S0 = 3.3
speedUpCoarseningP4S100 = 3.3
speedUpCoarseningP16S0 = 11
speedUpCoarseningP16S100 = 12.6
speedUpCoarseningP64S0 = 23.3
speedUpCoarseningP64S100 = 33.2
speedUpInitialPartitiongP4S0 = 2.7
speedUpInitialPartitiongP4S100 = 3.8
speedUpInitialPartitiongP16S0 = 3.3
speedUpInitialPartitiongP16S100 = 10.9
speedUpInitialPartitiongP64S0 = 4.4
speedUpInitialPartitiongP64S100 = 16.2
speedUpBatchUncontractionP4S0 = 3.9
speedUpBatchUncontractionP4S100 = 3.8
speedUpBatchUncontractionP16S0 = 10.8
speedUpBatchUncontractionP16S100 = 11.6
speedUpBatchUncontractionP64S0 = 20.8
speedUpBatchUncontractionP64S100 = 22.5
speedUpLabelPropagationP4S0 = 3.8
speedUpLabelPropagationP4S100 = 4.1
speedUpLabelPropagationP16S0 = 7.3
speedUpLabelPropagationP16S100 = 10.6
speedUpLabelPropagationP64S0 = 12.9
speedUpLabelPropagationP64S100 = 24.2
speedUpFMP4S0 = 3.9
speedUpFMP4S100 = 3.9
speedUpFMP16S0 = 10.6
speedUpFMP16S100 = 11.5
speedUpFMP64S0 = 14.8
speedUpFMP64S100 = 20.2
percentageInstancesGreaterTotalTimeS100 = 71.8
percentageInstancesGreaterCoarseningS100 = 31.5
percentageInstancesGreaterInitialPartitioningS100 = 7.8
percentageInstancesGreaterBatchUncontractionS100 = 34.4
percentageInstancesGreaterLabelPropagationS100 = 18.5
percentageInstancesGreaterFMS100 = 39
percentageInstancesSmallerTotalTimeS100 = 28.2
percentageInstancesSmallerCoarseningS100 = 68.5
percentageInstancesSmallerInitialPartitioningS100 = 92.2
percentageInstancesSmallerBatchUncontractionS100 = 65.6
percentageInstancesSmallerLabelPropagationS100 = 81.5
percentageInstancesSmallerFMS100 = 61
\end{filecontents*}
\DTLloaddb[noheader, keys={key,value}]{speedup}{experiments/scalability_experiments/speedup.dat}
\begin{filecontents*}{experiments/scalability_experiments/share.dat}
numBatchUncontractionInstancesWithShareSmaller10P1 = 32.1
numBatchUncontractionInstancesWithShareGreater10P1 = 67.9
numBatchUncontractionInstancesWithShareSmaller10P4 = 32.8
numBatchUncontractionInstancesWithShareGreater10P4 = 67.2
numBatchUncontractionInstancesWithShareSmaller10P16 = 28.9
numBatchUncontractionInstancesWithShareGreater10P16 = 71.1
numBatchUncontractionInstancesWithShareSmaller10P64 = 31.2
numBatchUncontractionInstancesWithShareGreater10P64 = 68.8
numBatchUncontractionInstancesWithShareSmaller25P1 = 63.3
numBatchUncontractionInstancesWithShareGreater25P1 = 36.7
numBatchUncontractionInstancesWithShareSmaller25P4 = 65.3
numBatchUncontractionInstancesWithShareGreater25P4 = 34.7
numBatchUncontractionInstancesWithShareSmaller25P16 = 64
numBatchUncontractionInstancesWithShareGreater25P16 = 36
numBatchUncontractionInstancesWithShareSmaller25P64 = 68.5
numBatchUncontractionInstancesWithShareGreater25P64 = 31.5
numCoarseningInstancesWithShareSmaller10P1 = 24.4
numCoarseningInstancesWithShareGreater10P1 = 75.6
numCoarseningInstancesWithShareSmaller10P4 = 22.4
numCoarseningInstancesWithShareGreater10P4 = 77.6
numCoarseningInstancesWithShareSmaller10P16 = 22.7
numCoarseningInstancesWithShareGreater10P16 = 77.3
numCoarseningInstancesWithShareSmaller10P64 = 26.3
numCoarseningInstancesWithShareGreater10P64 = 73.7
numCoarseningInstancesWithShareSmaller25P1 = 64.9
numCoarseningInstancesWithShareGreater25P1 = 35.1
numCoarseningInstancesWithShareSmaller25P4 = 52.9
numCoarseningInstancesWithShareGreater25P4 = 47.1
numCoarseningInstancesWithShareSmaller25P16 = 60.7
numCoarseningInstancesWithShareGreater25P16 = 39.3
numCoarseningInstancesWithShareSmaller25P64 = 70.8
numCoarseningInstancesWithShareGreater25P64 = 29.2
numInitialPartitioningInstancesWithShareSmaller10P1 = 86.7
numInitialPartitioningInstancesWithShareGreater10P1 = 13.3
numInitialPartitioningInstancesWithShareSmaller10P4 = 87
numInitialPartitioningInstancesWithShareGreater10P4 = 13
numInitialPartitioningInstancesWithShareSmaller10P16 = 85.4
numInitialPartitioningInstancesWithShareGreater10P16 = 14.6
numInitialPartitioningInstancesWithShareSmaller10P64 = 72.7
numInitialPartitioningInstancesWithShareGreater10P64 = 27.3
numInitialPartitioningInstancesWithShareSmaller25P1 = 98.7
numInitialPartitioningInstancesWithShareGreater25P1 = 1.3
numInitialPartitioningInstancesWithShareSmaller25P4 = 98.7
numInitialPartitioningInstancesWithShareGreater25P4 = 1.3
numInitialPartitioningInstancesWithShareSmaller25P16 = 98.4
numInitialPartitioningInstancesWithShareGreater25P16 = 1.6
numInitialPartitioningInstancesWithShareSmaller25P64 = 96.1
numInitialPartitioningInstancesWithShareGreater25P64 = 3.9
numLabelPropagationInstancesWithShareSmaller10P1 = 89.9
numLabelPropagationInstancesWithShareGreater10P1 = 10.1
numLabelPropagationInstancesWithShareSmaller10P4 = 93.2
numLabelPropagationInstancesWithShareGreater10P4 = 6.8
numLabelPropagationInstancesWithShareSmaller10P16 = 88.6
numLabelPropagationInstancesWithShareGreater10P16 = 11.4
numLabelPropagationInstancesWithShareSmaller10P64 = 91.6
numLabelPropagationInstancesWithShareGreater10P64 = 8.4
numLabelPropagationInstancesWithShareSmaller25P1 = 97.7
numLabelPropagationInstancesWithShareGreater25P1 = 2.3
numLabelPropagationInstancesWithShareSmaller25P4 = 97.4
numLabelPropagationInstancesWithShareGreater25P4 = 2.6
numLabelPropagationInstancesWithShareSmaller25P16 = 97.4
numLabelPropagationInstancesWithShareGreater25P16 = 2.6
numLabelPropagationInstancesWithShareSmaller25P64 = 99
numLabelPropagationInstancesWithShareGreater25P64 = 1
numFMInstancesWithShareSmaller10P1 = 33.4
numFMInstancesWithShareGreater10P1 = 66.6
numFMInstancesWithShareSmaller10P4 = 35.4
numFMInstancesWithShareGreater10P4 = 64.6
numFMInstancesWithShareSmaller10P16 = 31.5
numFMInstancesWithShareGreater10P16 = 68.5
numFMInstancesWithShareSmaller10P64 = 30.8
numFMInstancesWithShareGreater10P64 = 69.2
numFMInstancesWithShareSmaller25P1 = 56.5
numFMInstancesWithShareGreater25P1 = 43.5
numFMInstancesWithShareSmaller25P4 = 59.1
numFMInstancesWithShareGreater25P4 = 40.9
numFMInstancesWithShareSmaller25P16 = 59.1
numFMInstancesWithShareGreater25P16 = 40.9
numFMInstancesWithShareSmaller25P64 = 55.8
numFMInstancesWithShareGreater25P64 = 44.2
numCombinedInstancesWithShareGreater50P1 = 93.8
numCombinedInstancesWithShareGreater50P4 = 93.2
numCombinedInstancesWithShareGreater50P16 = 94.5
numCombinedInstancesWithShareGreater50P64 = 92.5
numCombinedInstancesWithShareGreater60P1 = 72.1
numCombinedInstancesWithShareGreater60P4 = 69.8
numCombinedInstancesWithShareGreater60P16 = 72.7
numCombinedInstancesWithShareGreater60P64 = 67.5
numCombinedInstancesWithShareGreater75P1 = 23.1
numCombinedInstancesWithShareGreater75P4 = 25.6
numCombinedInstancesWithShareGreater75P16 = 23.7
numCombinedInstancesWithShareGreater75P64 = 19.5
\end{filecontents*}
\DTLloaddb[noheader, keys={key,value}]{share}{experiments/scalability_experiments/share.dat}
\begin{filecontents*}{experiments/kahypar_benchmark_set/sequential.dat}
mtKaHyParStrongvsKaHyParHFCMtKaHyParStrong10Tau100 = 13.4
mtKaHyParStrongvsKaHyParHFCMtKaHyParStrong10TauInv100 = 86.6
mtKaHyParStrongvsKaHyParHFCMtKaHyParStrong10Tau110 = 87.4
mtKaHyParStrongvsKaHyParHFCMtKaHyParStrong10TauInv110 = 12.6
mtKaHyParStrongvsKaHyParHFCKaHyParHFCTau100 = 88.2
mtKaHyParStrongvsKaHyParHFCKaHyParHFCTauInv100 = 11.8
mtKaHyParStrongvsKaHyParHFCKaHyParHFCTau110 = 98.2
mtKaHyParStrongvsKaHyParHFCKaHyParHFCTauInv110 = 1.8
mtKaHyParStrongvsKaHyParCAMtKaHyParStrong10Tau100 = 40.4
mtKaHyParStrongvsKaHyParCAMtKaHyParStrong10TauInv100 = 59.6
mtKaHyParStrongvsKaHyParCAMtKaHyParStrong10Tau110 = 95.8
mtKaHyParStrongvsKaHyParCAMtKaHyParStrong10TauInv110 = 4.2
mtKaHyParStrongvsKaHyParCAKaHyParCATau100 = 61.3
mtKaHyParStrongvsKaHyParCAKaHyParCATauInv100 = 38.7
mtKaHyParStrongvsKaHyParCAKaHyParCATau110 = 97.3
mtKaHyParStrongvsKaHyParCAKaHyParCATauInv110 = 2.7
mtKaHyParStrongvsmtKaHyParFastMtKaHyParStrong10Tau100 = 87.8
mtKaHyParStrongvsmtKaHyParFastMtKaHyParStrong10TauInv100 = 12.2
mtKaHyParStrongvsmtKaHyParFastMtKaHyParStrong10Tau110 = 99.4
mtKaHyParStrongvsmtKaHyParFastMtKaHyParStrong10TauInv110 = 0.6
mtKaHyParStrongvsmtKaHyParFastMtKaHyParFast10Tau100 = 13.6
mtKaHyParStrongvsmtKaHyParFastMtKaHyParFast10TauInv100 = 86.4
mtKaHyParStrongvsmtKaHyParFastMtKaHyParFast10Tau110 = 95.9
mtKaHyParStrongvsmtKaHyParFastMtKaHyParFast10TauInv110 = 4.1
mtKaHyParStrongvshMetisRMtKaHyParStrong10Tau100 = 52.1
mtKaHyParStrongvshMetisRMtKaHyParStrong10TauInv100 = 47.9
mtKaHyParStrongvshMetisRMtKaHyParStrong10Tau110 = 89.9
mtKaHyParStrongvshMetisRMtKaHyParStrong10TauInv110 = 10.1
mtKaHyParStrongvshMetisRhMetisRTau100 = 49.2
mtKaHyParStrongvshMetisRhMetisRTauInv100 = 50.8
mtKaHyParStrongvshMetisRhMetisRTau110 = 82
mtKaHyParStrongvshMetisRhMetisRTauInv110 = 18
mtKaHyParStrongvsPaToHDMtKaHyParStrong10Tau100 = 88.2
mtKaHyParStrongvsPaToHDMtKaHyParStrong10TauInv100 = 11.8
mtKaHyParStrongvsPaToHDMtKaHyParStrong10Tau110 = 97
mtKaHyParStrongvsPaToHDMtKaHyParStrong10TauInv110 = 3
mtKaHyParStrongvsPaToHDPaToHDTau100 = 12.7
mtKaHyParStrongvsPaToHDPaToHDTauInv100 = 87.3
mtKaHyParStrongvsPaToHDPaToHDTau110 = 56.5
mtKaHyParStrongvsPaToHDPaToHDTauInv110 = 43.5
mtKaHyParStrongvsPaToHQMtKaHyParStrong10Tau100 = 70.1
mtKaHyParStrongvsPaToHQMtKaHyParStrong10TauInv100 = 29.9
mtKaHyParStrongvsPaToHQMtKaHyParStrong10Tau110 = 93.9
mtKaHyParStrongvsPaToHQMtKaHyParStrong10TauInv110 = 6.1
mtKaHyParStrongvsPaToHQPaToHQTau100 = 31.1
mtKaHyParStrongvsPaToHQPaToHQTauInv100 = 68.9
mtKaHyParStrongvsPaToHQPaToHQTau110 = 81.1
mtKaHyParStrongvsPaToHQPaToHQTauInv110 = 18.9
mtKaHyParFastvsPaToHDMtKaHyParFast10Tau100 = 81.7
mtKaHyParFastvsPaToHDMtKaHyParFast10TauInv100 = 18.3
mtKaHyParFastvsPaToHDMtKaHyParFast10Tau110 = 95.7
mtKaHyParFastvsPaToHDMtKaHyParFast10TauInv110 = 4.3
mtKaHyParFastvsPaToHDPaToHDTau100 = 19.1
mtKaHyParFastvsPaToHDPaToHDTauInv100 = 80.9
mtKaHyParFastvsPaToHDPaToHDTau110 = 66.2
mtKaHyParFastvsPaToHDPaToHDTauInv110 = 33.8
mtKaHyParFastvsPaToHQMtKaHyParFast10Tau100 = 57.4
mtKaHyParFastvsPaToHQMtKaHyParFast10TauInv100 = 42.6
mtKaHyParFastvsPaToHQMtKaHyParFast10Tau110 = 90.5
mtKaHyParFastvsPaToHQMtKaHyParFast10TauInv110 = 9.5
mtKaHyParFastvsPaToHQPaToHQTau100 = 43.8
mtKaHyParFastvsPaToHQPaToHQTauInv100 = 56.2
mtKaHyParFastvsPaToHQPaToHQTau110 = 84.8
mtKaHyParFastvsPaToHQPaToHQTauInv110 = 15.2
mtKaHyParStrongVsKaHyParCAMtKaHyParStrongV10Tau100 = 52.5
mtKaHyParStrongVsKaHyParCAMtKaHyParStrongV10TauInv100 = 47.5
mtKaHyParStrongVsKaHyParCAMtKaHyParStrongV10Tau110 = 96.9
mtKaHyParStrongVsKaHyParCAMtKaHyParStrongV10TauInv110 = 3.1
mtKaHyParStrongVsKaHyParCAKaHyParCATau100 = 49.2
mtKaHyParStrongVsKaHyParCAKaHyParCATauInv100 = 50.8
mtKaHyParStrongVsKaHyParCAKaHyParCATau110 = 96.5
mtKaHyParStrongVsKaHyParCAKaHyParCATauInv110 = 3.5
MtKaHyParStrong10vsMtKaHyParFast10GmeanTimeRatio = 3.4
MtKaHyParStrong10vsMtKaHyParStrongV10GmeanTimeRatio = 0.7
MtKaHyParStrong10vsMtKaHyParFastAlenex10GmeanTimeRatio = 2.1
MtKaHyParStrong10vsKaHyParCAGmeanTimeRatio = 0.1
MtKaHyParStrong10vsKaHyParHFCGmeanTimeRatio = 0.1
MtKaHyParStrong10vsPaToHDGmeanTimeRatio = 2.7
MtKaHyParFast10vsMtKaHyParStrong10GmeanTimeRatio = 0.3
MtKaHyParFast10vsMtKaHyParStrongV10GmeanTimeRatio = 0.2
MtKaHyParFast10vsMtKaHyParFastAlenex10GmeanTimeRatio = 0.6
MtKaHyParFast10vsKaHyParCAGmeanTimeRatio = 0
MtKaHyParFast10vsKaHyParHFCGmeanTimeRatio = 0
MtKaHyParFast10vsPaToHDGmeanTimeRatio = 0.8
MtKaHyParStrongV10vsMtKaHyParStrong10GmeanTimeRatio = 1.4
MtKaHyParStrongV10vsMtKaHyParFast10GmeanTimeRatio = 4.7
MtKaHyParStrongV10vsMtKaHyParFastAlenex10GmeanTimeRatio = 3
MtKaHyParStrongV10vsKaHyParCAGmeanTimeRatio = 0.2
MtKaHyParStrongV10vsKaHyParHFCGmeanTimeRatio = 0.1
MtKaHyParStrongV10vsPaToHDGmeanTimeRatio = 3.8
MtKaHyParFastAlenex10vsMtKaHyParStrong10GmeanTimeRatio = 0.5
MtKaHyParFastAlenex10vsMtKaHyParFast10GmeanTimeRatio = 1.6
MtKaHyParFastAlenex10vsMtKaHyParStrongV10GmeanTimeRatio = 0.3
MtKaHyParFastAlenex10vsKaHyParCAGmeanTimeRatio = 0.1
MtKaHyParFastAlenex10vsKaHyParHFCGmeanTimeRatio = 0
MtKaHyParFastAlenex10vsPaToHDGmeanTimeRatio = 1.3
KaHyParCAvsMtKaHyParStrong10GmeanTimeRatio = 8.8
KaHyParCAvsMtKaHyParFast10GmeanTimeRatio = 29.7
KaHyParCAvsMtKaHyParStrongV10GmeanTimeRatio = 6.3
KaHyParCAvsMtKaHyParFastAlenex10GmeanTimeRatio = 18.7
KaHyParCAvsKaHyParHFCGmeanTimeRatio = 0.6
KaHyParCAvsPaToHDGmeanTimeRatio = 24
KaHyParHFCvsMtKaHyParStrong10GmeanTimeRatio = 15.4
KaHyParHFCvsMtKaHyParFast10GmeanTimeRatio = 51.6
KaHyParHFCvsMtKaHyParStrongV10GmeanTimeRatio = 11
KaHyParHFCvsMtKaHyParFastAlenex10GmeanTimeRatio = 32.6
KaHyParHFCvsKaHyParCAGmeanTimeRatio = 1.7
KaHyParHFCvsPaToHDGmeanTimeRatio = 41.7
PaToHDvsMtKaHyParStrong10GmeanTimeRatio = 0.4
PaToHDvsMtKaHyParFast10GmeanTimeRatio = 1.2
PaToHDvsMtKaHyParStrongV10GmeanTimeRatio = 0.3
PaToHDvsMtKaHyParFastAlenex10GmeanTimeRatio = 0.8
PaToHDvsKaHyParCAGmeanTimeRatio = 0
PaToHDvsKaHyParHFCGmeanTimeRatio = 0
FasterThanMtKaHyParFast10On = 0.2
SlowerThanMtKaHyParFast10On = 99.8
FasterThanMtKaHyParFast10On25PercentByFactor = 0.4
FasterThanMtKaHyParFast10On50PercentByFactor = 0.3
FasterThanMtKaHyParFast10On75PercentByFactor = 0.2
FasterThanMtKaHyParStrongV10On = 99.7
SlowerThanMtKaHyParStrongV10On = 0.3
FasterThanMtKaHyParStrongV10On25PercentByFactor = 1.6
FasterThanMtKaHyParStrongV10On50PercentByFactor = 1.4
FasterThanMtKaHyParStrongV10On75PercentByFactor = 1.2
FasterThanKaHyParCAOn = 99.9
SlowerThanKaHyParCAOn = 0.1
FasterThanKaHyParCAOn25PercentByFactor = 12.7
FasterThanKaHyParCAOn50PercentByFactor = 7.5
FasterThanKaHyParCAOn75PercentByFactor = 5.2
FasterThanKaHyParHFCOn = 100
SlowerThanKaHyParHFCOn = 0
FasterThanKaHyParHFCOn25PercentByFactor = 26.8
FasterThanKaHyParHFCOn50PercentByFactor = 13
FasterThanKaHyParHFCOn75PercentByFactor = 7.4
gmeanTimeMtKaHyParFast10 = 0.95
gmeanTimePaToHD = 1.17
gmeanTimeMtKaHyParStrong10 = 3.19
gmeanTimePaToHQ = 5.86
gmeanTimeKaHyParCA = 28.14
gmeanTimeKaHyParHFC = 48.98
\end{filecontents*}
\DTLloaddb[noheader, keys={key,value}]{sequential}{experiments/kahypar_benchmark_set/sequential.dat}
\begin{filecontents*}{experiments/big_benchmark_set/parallel.dat}
mtKaHyParFastvsPaToHDMtKaHyParFast64Tau100 = 79.5
mtKaHyParFastvsPaToHDMtKaHyParFast64TauInv100 = 20.5
mtKaHyParFastvsPaToHDMtKaHyParFast64Tau110 = 95.7
mtKaHyParFastvsPaToHDMtKaHyParFast64TauInv110 = 4.3
mtKaHyParFastvsPaToHDPaToHDTau100 = 20.7
mtKaHyParFastvsPaToHDPaToHDTauInv100 = 79.3
mtKaHyParFastvsPaToHDPaToHDTau110 = 59.8
mtKaHyParFastvsPaToHDPaToHDTauInv110 = 40.2
mtKaHyParFastvsBiPartMtKaHyParFast64Tau100 = 98.1
mtKaHyParFastvsBiPartMtKaHyParFast64TauInv100 = 1.9
mtKaHyParFastvsBiPartMtKaHyParFast64Tau110 = 99.5
mtKaHyParFastvsBiPartMtKaHyParFast64TauInv110 = 0.5
mtKaHyParFastvsBiPartBiPart64Tau100 = 1.9
mtKaHyParFastvsBiPartBiPart64TauInv100 = 98.1
mtKaHyParFastvsBiPartBiPart64Tau110 = 6.6
mtKaHyParFastvsBiPartBiPart64TauInv110 = 93.4
mtKaHyParFastvsZoltanMtKaHyParFast64Tau100 = 94.7
mtKaHyParFastvsZoltanMtKaHyParFast64TauInv100 = 5.3
mtKaHyParFastvsZoltanMtKaHyParFast64Tau110 = 98.9
mtKaHyParFastvsZoltanMtKaHyParFast64TauInv110 = 1.1
mtKaHyParFastvsZoltanZoltan64Tau100 = 6.1
mtKaHyParFastvsZoltanZoltan64TauInv100 = 93.9
mtKaHyParFastvsZoltanZoltan64Tau110 = 21
mtKaHyParFastvsZoltanZoltan64TauInv110 = 79
mtKaHyParFastvsHypeMtKaHyParFast64Tau100 = 100
mtKaHyParFastvsHypeMtKaHyParFast64TauInv100 = 0
mtKaHyParFastvsHypeHypeTau100 = 0
mtKaHyParFastvsHypeHypeTauInv100 = 100
mtKaHyParFastvsHypeHypeTau110 = 0
mtKaHyParFastvsHypeHypeTauInv110 = 100
mtKaHyParFastvsMtKahyparStrongMtKaHyParFast64Tau100 = 19.7
mtKaHyParFastvsMtKahyparStrongMtKaHyParFast64TauInv100 = 80.3
mtKaHyParFastvsMtKahyparStrongMtKaHyParFast64Tau110 = 94.7
mtKaHyParFastvsMtKahyparStrongMtKaHyParFast64TauInv110 = 5.3
mtKaHyParFastvsMtKahyparStrongMtKaHyParStrong64Tau100 = 81.4
mtKaHyParFastvsMtKahyparStrongMtKaHyParStrong64TauInv100 = 18.6
mtKaHyParFastvsMtKahyparStrongMtKaHyParStrong64Tau110 = 96.8
mtKaHyParFastvsMtKahyparStrongMtKaHyParStrong64TauInv110 = 3.2
mtKaHyParStrongvsPaToHDMtKaHyParStrong64Tau100 = 83.8
mtKaHyParStrongvsPaToHDMtKaHyParStrong64TauInv100 = 16.2
mtKaHyParStrongvsPaToHDMtKaHyParStrong64Tau110 = 96.3
mtKaHyParStrongvsPaToHDMtKaHyParStrong64TauInv110 = 3.7
mtKaHyParStrongvsPaToHDPaToHDTau100 = 16
mtKaHyParStrongvsPaToHDPaToHDTauInv100 = 84
mtKaHyParStrongvsPaToHDPaToHDTau110 = 52.4
mtKaHyParStrongvsPaToHDPaToHDTauInv110 = 47.6
mtKaHyParStrongSvsBiPartMtKaHyParStrong64Tau100 = 96.3
mtKaHyParStrongSvsBiPartMtKaHyParStrong64TauInv100 = 3.7
mtKaHyParStrongSvsBiPartMtKaHyParStrong64Tau110 = 97.3
mtKaHyParStrongSvsBiPartMtKaHyParStrong64TauInv110 = 2.7
mtKaHyParStrongSvsBiPartBiPart64Tau100 = 3.7
mtKaHyParStrongSvsBiPartBiPart64TauInv100 = 96.3
mtKaHyParStrongSvsBiPartBiPart64Tau110 = 6.9
mtKaHyParStrongSvsBiPartBiPart64TauInv110 = 93.1
mtKaHyParStrongvsZoltanMtKaHyParStrong64Tau100 = 94.7
mtKaHyParStrongvsZoltanMtKaHyParStrong64TauInv100 = 5.3
mtKaHyParStrongvsZoltanMtKaHyParStrong64Tau110 = 96.8
mtKaHyParStrongvsZoltanMtKaHyParStrong64TauInv110 = 3.2
mtKaHyParStrongvsZoltanZoltan64Tau100 = 6.1
mtKaHyParStrongvsZoltanZoltan64TauInv100 = 93.9
mtKaHyParStrongvsZoltanZoltan64Tau110 = 17.6
mtKaHyParStrongvsZoltanZoltan64TauInv110 = 82.4
mtKaHyParStrongvsHypeMtKaHyParStrong64Tau100 = 98.4
mtKaHyParStrongvsHypeMtKaHyParStrong64TauInv100 = 1.6
mtKaHyParStrongvsHypeMtKaHyParStrong64Tau110 = 98.4
mtKaHyParStrongvsHypeMtKaHyParStrong64TauInv110 = 1.6
mtKaHyParStrongvsHypeHypeTau100 = 1.6
mtKaHyParStrongvsHypeHypeTauInv100 = 98.4
mtKaHyParStrongvsHypeHypeTau110 = 1.6
mtKaHyParStrongvsHypeHypeTauInv110 = 98.4
MtKaHyParStrong64vsMtKaHyParFast64GmeanTimeRatio = 6.3
MtKaHyParStrong64vsPaToHDGmeanTimeRatio = 0.6
MtKaHyParStrong64vsBiPart64GmeanTimeRatio = 1.1
MtKaHyParStrong64vsZoltan64GmeanTimeRatio = 2.4
MtKaHyParStrong64vsHypeGmeanTimeRatio = 1.2
MtKaHyParFast64vsMtKaHyParStrong64GmeanTimeRatio = 0.2
MtKaHyParFast64vsPaToHDGmeanTimeRatio = 0.1
MtKaHyParFast64vsBiPart64GmeanTimeRatio = 0.2
MtKaHyParFast64vsZoltan64GmeanTimeRatio = 0.4
MtKaHyParFast64vsHypeGmeanTimeRatio = 0.2
PaToHDvsMtKaHyParStrong64GmeanTimeRatio = 1.7
PaToHDvsMtKaHyParFast64GmeanTimeRatio = 10.5
PaToHDvsBiPart64GmeanTimeRatio = 1.8
PaToHDvsZoltan64GmeanTimeRatio = 4.1
PaToHDvsHypeGmeanTimeRatio = 2
BiPart64vsMtKaHyParStrong64GmeanTimeRatio = 1
BiPart64vsMtKaHyParFast64GmeanTimeRatio = 6
BiPart64vsPaToHDGmeanTimeRatio = 0.6
BiPart64vsZoltan64GmeanTimeRatio = 2.3
BiPart64vsHypeGmeanTimeRatio = 1.1
Zoltan64vsMtKaHyParStrong64GmeanTimeRatio = 0.4
Zoltan64vsMtKaHyParFast64GmeanTimeRatio = 2.6
Zoltan64vsPaToHDGmeanTimeRatio = 0.2
Zoltan64vsBiPart64GmeanTimeRatio = 0.4
Zoltan64vsHypeGmeanTimeRatio = 0.5
HypevsMtKaHyParStrong64GmeanTimeRatio = 0.8
HypevsMtKaHyParFast64GmeanTimeRatio = 5.2
HypevsPaToHDGmeanTimeRatio = 0.5
HypevsBiPart64GmeanTimeRatio = 0.9
HypevsZoltan64GmeanTimeRatio = 2
FasterThanMtKaHyParFast64On = 0
SlowerThanMtKaHyParFast64On = 100
FasterThanMtKaHyParFast64On25PercentByFactor = 0.3
FasterThanMtKaHyParFast64On50PercentByFactor = 0.2
FasterThanMtKaHyParFast64On75PercentByFactor = 0.1
FasterThanPaToHDOn = 66.5
SlowerThanPaToHDOn = 33.5
FasterThanPaToHDOn25PercentByFactor = 4.5
FasterThanPaToHDOn50PercentByFactor = 1.8
FasterThanPaToHDOn75PercentByFactor = 0.7
FasterThanBiPart64On = 52.7
SlowerThanBiPart64On = 47.3
FasterThanBiPart64On25PercentByFactor = 2.4
FasterThanBiPart64On50PercentByFactor = 1.1
FasterThanBiPart64On75PercentByFactor = 0.4
FasterThanZoltan64On = 34
SlowerThanZoltan64On = 66
FasterThanZoltan64On25PercentByFactor = 1.2
FasterThanZoltan64On50PercentByFactor = 0.6
FasterThanZoltan64On75PercentByFactor = 0.2
FasterThanHypeOn = 50.8
SlowerThanHypeOn = 49.2
FasterThanHypeOn25PercentByFactor = 2.9
FasterThanHypeOn50PercentByFactor = 1
FasterThanHypeOn75PercentByFactor = 0.2
gmeanTimeMtKaHyParFast64 = 4.89
gmeanTimeZoltan64 = 12.63
gmeanTimeBiPart64 = 29.19
gmeanTimeMtKaHyParStrong64 = 30.7
gmeanTimePaToHD = 51.2
gmeanTimeHype = 25.56
\end{filecontents*}
\DTLloaddb[noheader, keys={key,value}]{parallel}{experiments/big_benchmark_set/parallel.dat}
\begin{filecontents*}{experiments/shp_comparison/shp.dat}
MtKaHyParFastemailEnronK2 = 5220
MtKaHyParFastemailEnronK8 = 17925
MtKaHyParFastemailEnronK32 = 34206
MtKaHyParFastemailEnronK128 = 58607
MtKaHyParFastemailEnronK2Time = 0.25
MtKaHyParFastemailEnronK8Time = 0.48
MtKaHyParFastemailEnronK32Time = 0.88
MtKaHyParFastemailEnronK128Time = 1.26
MtKaHyParFastsocEpinionsK2 = 902
MtKaHyParFastsocEpinionsK8 = 26843
MtKaHyParFastsocEpinionsK32 = 62397
MtKaHyParFastsocEpinionsK128 = 118675
MtKaHyParFastsocEpinionsK2Time = 0.33
MtKaHyParFastsocEpinionsK8Time = 0.71
MtKaHyParFastsocEpinionsK32Time = 1.16
MtKaHyParFastsocEpinionsK128Time = 1.9
MtKaHyParFastwebStanfordK2 = 43
MtKaHyParFastwebStanfordK8 = 10323
MtKaHyParFastwebStanfordK32 = 35225
MtKaHyParFastwebStanfordK128 = 67899
MtKaHyParFastwebStanfordK2Time = 0.45
MtKaHyParFastwebStanfordK8Time = 0.54
MtKaHyParFastwebStanfordK32Time = 0.83
MtKaHyParFastwebStanfordK128Time = 1.39
MtKaHyParFastwebBerkStanK2 = 349
MtKaHyParFastwebBerkStanK8 = 19828
MtKaHyParFastwebBerkStanK32 = 90102
MtKaHyParFastwebBerkStanK128 = 181492
MtKaHyParFastwebBerkStanK2Time = 0.78
MtKaHyParFastwebBerkStanK8Time = 1.01
MtKaHyParFastwebBerkStanK32Time = 1.57
MtKaHyParFastwebBerkStanK128Time = 2.48
MtKaHyParFastsocPokecK2 = 498597
MtKaHyParFastsocPokecK8 = 1870665
MtKaHyParFastsocPokecK32 = 3576791
MtKaHyParFastsocPokecK128 = 5502354
MtKaHyParFastsocPokecK2Time = 18.37
MtKaHyParFastsocPokecK8Time = 39.77
MtKaHyParFastsocPokecK32Time = 57.19
MtKaHyParFastsocPokecK128Time = 87.8
MtKaHyParFastsocLiveJournalK2 = 751758
MtKaHyParFastsocLiveJournalK8 = 2684767
MtKaHyParFastsocLiveJournalK32 = 5230242
MtKaHyParFastsocLiveJournalK128 = 8268903
MtKaHyParFastsocLiveJournalK2Time = 33.57
MtKaHyParFastsocLiveJournalK8Time = 74.11
MtKaHyParFastsocLiveJournalK32Time = 104.83
MtKaHyParFastsocLiveJournalK128Time = 169.04
MtKaHyParStrongemailEnronK2 = 4473
MtKaHyParStrongemailEnronK8 = 16476
MtKaHyParStrongemailEnronK32 = 33480
MtKaHyParStrongemailEnronK128 = 52321
MtKaHyParStrongemailEnronK2Time = 0.62
MtKaHyParStrongemailEnronK8Time = 1.48
MtKaHyParStrongemailEnronK32Time = 4.08
MtKaHyParStrongemailEnronK128Time = 10.63
MtKaHyParStrongsocEpinionsK2 = 750
MtKaHyParStrongsocEpinionsK8 = 24261
MtKaHyParStrongsocEpinionsK32 = 55945
MtKaHyParStrongsocEpinionsK128 = 92989
MtKaHyParStrongsocEpinionsK2Time = 1.35
MtKaHyParStrongsocEpinionsK8Time = 3.18
MtKaHyParStrongsocEpinionsK32Time = 9.27
MtKaHyParStrongsocEpinionsK128Time = 23.93
MtKaHyParStrongwebStanfordK2 = 50
MtKaHyParStrongwebStanfordK8 = 9665
MtKaHyParStrongwebStanfordK32 = 33949
MtKaHyParStrongwebStanfordK128 = 66393
MtKaHyParStrongwebStanfordK2Time = 7.74
MtKaHyParStrongwebStanfordK8Time = 7.29
MtKaHyParStrongwebStanfordK32Time = 8.79
MtKaHyParStrongwebStanfordK128Time = 15.71
MtKaHyParStrongwebBerkStanK2 = 774
MtKaHyParStrongwebBerkStanK8 = 16185
MtKaHyParStrongwebBerkStanK32 = 88997
MtKaHyParStrongwebBerkStanK128 = 179311
MtKaHyParStrongwebBerkStanK2Time = 23.39
MtKaHyParStrongwebBerkStanK8Time = 18.44
MtKaHyParStrongwebBerkStanK32Time = 27.82
MtKaHyParStrongwebBerkStanK128Time = 98.71
MtKaHyParStrongsocPokecK2 = 487483
MtKaHyParStrongsocPokecK8 = 1880231
MtKaHyParStrongsocPokecK32 = 3529972
MtKaHyParStrongsocPokecK128 = 5473653
MtKaHyParStrongsocPokecK2Time = 210.06
MtKaHyParStrongsocPokecK8Time = 274.96
MtKaHyParStrongsocPokecK32Time = 605.74
MtKaHyParStrongsocPokecK128Time = 1881.49
MtKaHyParStrongsocLiveJournalK2 = 740468
MtKaHyParStrongsocLiveJournalK8 = 2638341
MtKaHyParStrongsocLiveJournalK32 = 5118153
MtKaHyParStrongsocLiveJournalK128 = 7996864
MtKaHyParStrongsocLiveJournalK2Time = 378.97
MtKaHyParStrongsocLiveJournalK8Time = 568.13
MtKaHyParStrongsocLiveJournalK32Time = 1507.88
MtKaHyParStrongsocLiveJournalK128Time = 23816.26
SHP2emailEnronK2 = 3313
SHP2emailEnronK8 = 19875
SHP2emailEnronK32 = 39241
SHP2emailEnronK128 = 65996
SHP2emailEnronK2Time = -
SHP2emailEnronK8Time = -
SHP2emailEnronK32Time = -
SHP2emailEnronK128Time = -
SHP2socEpinionsK2 = 1246
SHP2socEpinionsK8 = 21804
SHP2socEpinionsK32 = 53888
SHP2socEpinionsK128 = 95627
SHP2socEpinionsK2Time = -
SHP2socEpinionsK8Time = -
SHP2socEpinionsK32Time = -
SHP2socEpinionsK128Time = -
SHP2webStanfordK2 = 22779
SHP2webStanfordK8 = 60743
SHP2webStanfordK32 = 101239
SHP2webStanfordK128 = 149327
SHP2webStanfordK2Time = -
SHP2webStanfordK8Time = -
SHP2webStanfordK32Time = -
SHP2webStanfordK128Time = -
SHP2webBerkStanK2 = 67048
SHP2webBerkStanK8 = 170668
SHP2webBerkStanK32 = 274287
SHP2webBerkStanK128 = 420574
SHP2webBerkStanK2Time = -
SHP2webBerkStanK8Time = -
SHP2webBerkStanK32Time = -
SHP2webBerkStanK128Time = -
SHP2socPokecK2 = 625731
SHP2socPokecK8 = 2273064
SHP2socPokecK32 = 4175797
SHP2socPokecK128 = 6142380
SHP2socPokecK2Time = -
SHP2socPokecK8Time = -
SHP2socPokecK32Time = 108
SHP2socPokecK128Time = -
SHP2socLiveJournalK2 = 1051618
SHP2socLiveJournalK8 = 3969011
SHP2socLiveJournalK32 = 7768406
SHP2socLiveJournalK128 = 11805263
SHP2socLiveJournalK2Time = -
SHP2socLiveJournalK8Time = -
SHP2socLiveJournalK32Time = 144
SHP2socLiveJournalK128Time = -
SHPKemailEnronK2 = 3822
SHPKemailEnronK8 = 17837
SHPKemailEnronK32 = 33635
SHPKemailEnronK128 = 60900
SHPKemailEnronK2Time = -
SHPKemailEnronK8Time = -
SHPKemailEnronK32Time = -
SHPKemailEnronK128Time = -
SHPKsocEpinionsK2 = 1869
SHPKsocEpinionsK8 = 20247
SHPKsocEpinionsK32 = 47658
SHPKsocEpinionsK128 = 90332
SHPKsocEpinionsK2Time = -
SHPKsocEpinionsK8Time = -
SHPKsocEpinionsK32Time = -
SHPKsocEpinionsK128Time = -
SHPKwebStanfordK2 = 20248
SHPKwebStanfordK8 = 43026
SHPKwebStanfordK32 = 75929
SHPKwebStanfordK128 = 116425
SHPKwebStanfordK2Time = -
SHPKwebStanfordK8Time = -
SHPKwebStanfordK32Time = -
SHPKwebStanfordK128Time = -
SHPKwebBerkStanK2 = 60953
SHPKwebBerkStanK8 = 146286
SHPKwebBerkStanK32 = 195049
SHPKwebBerkStanK128 = 304764
SHPKwebBerkStanK2Time = -
SHPKwebBerkStanK8Time = -
SHPKwebBerkStanK32Time = -
SHPKwebBerkStanK128Time = -
SHPKsocPokecK2 = 549111
SHPKsocPokecK8 = 2094283
SHPKsocPokecK32 = 3920396
SHPKsocPokecK128 = 5772049
SHPKsocPokecK2Time = -
SHPKsocPokecK8Time = -
SHPKsocPokecK32Time = 138
SHPKsocPokecK128Time = -
SHPKsocLiveJournalK2 = 1255157
SHPKsocLiveJournalK8 = 3595856
SHPKsocLiveJournalK32 = 6275786
SHPKsocLiveJournalK128 = 9566334
SHPKsocLiveJournalK2Time = -
SHPKsocLiveJournalK8Time = -
SHPKsocLiveJournalK32Time = 276
SHPKsocLiveJournalK128Time = -
\end{filecontents*}
\DTLloaddb[noheader, keys={key,value}]{shp}{experiments/shp_comparison/shp.dat}
\begin{filecontents*}{experiments/maximum_batch_size/maximum_batch_size.dat}
gmeanbmax1 = 123.08
gmeanbmax100 = 15.51
gmeanbmax200 = 11.69
gmeanbmax1000 = 9.42
gmeanbmax10000 = 9.55
\end{filecontents*}
\DTLloaddb[noheader, keys={key,value}]{batch_size}{experiments/maximum_batch_size/maximum_batch_size.dat}

The proposed $n$-level approach is integrated in the hypergraph partitioning framework \Partitioner{Mt-KaHyPar}\footnote{\mtkahypar~is available from \url{https://github.com/kahypar/mt-kahypar}},
which is implemented in \texttt{C++17}, parallelized using the TBB library~\cite{TBB}, and compiled using \gpp{9.2} with the flags \texttt{-O3 -mtune=native -march=native}.
We refer to our new parallel $n$-level partitioner as \Partitioner{\mtkahyparnew} (for quality setting) and to the previously published parallel partitioner as \Partitioner{\mtkahyparold} (for default setting)~\cite{MT-KAHYPAR}.
For parallel partitioners we add a suffix to their name to indicate the number of threads used, e.g. \ExpAlgo{\mtkahyparnew}{64} for 64 threads.
We omit the suffix for sequential partitioners.


\subparagraph*{Instances.}

All instances of the benchmark sets used in the experimental evaluation are derived from four sources encompassing three application domains: the ISPD98 VLSI Circuit Benchmark Suite~\cite{ISPD98}, the DAC 2012 Routability-Driven Placement Contest~\cite{DAC}, the SuiteSparse Matrix Collection~\cite{SPM}, and the 2014 SAT Competition~\cite{SAT14}. VLSI instances are transformed into hypergraphs by converting the netlist of each circuit into a set of hyperedges.
Sparse matrices are translated to hypergraphs using the row-net model~\cite{PATOH} and SAT instances to three different hypergraph representations:
\emph{literal}, \emph{primal}, and \emph{dual}~\cite{MANN-PAPA14, PAPA-MARKOV} (see~\cite{KAHYPAR-CA} for more details).
All hypergraphs have unit vertex and net weights.

For comparison with sequential partitioners, we use the benchmark set of Heuer and Schlag~\cite{KAHYPAR-CA}, which contains $488$ hypergraphs (referred to as set A).
To measure speedups and to compare our implementation with other parallel partitioners,
we use a benchmark set composed of $94$ large hypergraphs (referred to as set B) that was initially assembled to evaluate
\Partitioner{\mtkahyparold}~\cite{MT-KAHYPAR}.
Figure~\ref{fig:benchmark_set} in Appendix~\ref{appendix:benchmark_stats} shows that the hypergraphs of set B are more
than an order of magnitude larger than those of set A.\footnote{The benchmark sets and detailed statistics of their properties are publicly available from \url{http://algo2.iti.kit.edu/heuer/nlevel/}.}


\subparagraph*{Algorithms.}

On set A, we compare \Partitioner{\mtkahyparold} and \Partitioner{\mtkahyparnew} with the following sequential hypergraph partitioners:
\Partitioner{KaHyPar-CA}~\cite{KAHYPAR-CA} ($n$-level partitioner with similar algorithmic components as \SeqAlgo{\mtkahyparnew}), \kahyparhfc~(the latest and strongest version, which extends
\SeqAlgo{KaHyPar-CA} with flow-based refinement~\cite{KAHYPAR-MF, REBAHFC}), as well as the default (\Partitioner{PaToH-D}), and quality preset (\Partitioner{PaToH-Q})
of \Partitioner{PaToH} $3.3$~\cite{PATOH}.
We do not consider \Partitioner{hMetis}~\cite{HMETIS-K, HMETIS} here, as \Partitioner{KaHyPar-CA} has already been shown to
outperform both the direct $k$-way and the recursive bipartitioning variant of \Partitioner{hMetis}, while also being
faster on average~\cite{KAHYPAR-CA}.
Similarly \Partitioner{PaToH-D} outperforms \Partitioner{Mondriaan}~\cite{MONDRIAAN} and \Partitioner{HYPE}~\cite{TheHypeIsOver} on set A.

On set B, we compare \Partitioner{\mtkahyparold} and \Partitioner{\mtkahyparnew} with the
distributed hypergraph partitioner \SeqAlgo{Zoltan} $3.83$~\cite{ZOLTAN},
the deterministic shared-memory hypergraph partitioner \Partitioner{BiPart}\footnote{\url{https://github.com/IntelligentSoftwareSystems/Galois/tree/release-6.0/lonestar/analytics/cpu/bipart}}~\cite{BIPART},
the sequential non-multilevel hypergraph partitioner \Partitioner{Hype}\footnote{\url{https://github.com/mayerrn/HYPE}}~\cite{TheHypeIsOver},
and \Partitioner{PaToH-D}.
While \Partitioner{Hype} and \Partitioner{PaToH-D} are fast enough to conduct the experiments on set B in a reasonable time frame,
this is not the case for \Partitioner{PaToH-Q} or any of the other sequential algorithms.

We could not include \Partitioner{Parkway}~\cite{PARKWAY-2} because the version available on GitHub either hangs infinitely or crashes with a segmentation fault\footnote{\url{https://github.com/parkway-partitioner/parkway}}. However, previous work showed
that its performance is comparable to Zoltan~\cite{ZOLTAN}.
The publicly available version of \Partitioner{SHP}~\cite{SHP} does not compile successfully, because package downloads during compilation time out\footnote{\url{https://issues.apache.org/jira/browse/GIRAPH-1131}}.
We therefore compare only with results reported in the paper~\cite{SHP}.
Additionally, the authors report that \Partitioner{SHP} computes worse partitions than \Partitioner{Zoltan}.

\subparagraph*{System and Methodology.}

We use two different machines.
Machines of type A are nodes of a cluster with Intel Xeon Gold 6230 processors ($2$ sockets with $20$ cores each) running at $2.1$ GHz with $96$GB RAM.
These machines are used for the comparison with sequential partitioners on set A, as these experiments took roughly 3.5 CPU years in total.
For these experiments, we use $k \in \{2,4,8,16,32,64,128\}$, $\varepsilon = 0.03$, ten different seeds and a time limit of eight hours.
In order to resemble workload scenarios on commodity multi-core machines, the parallel algorithms use $10$ threads for the experiments on set A.
Machine B is an AMD EPYC Rome 7702P ($1$ socket with $64$ cores) running at $2.0$--$3.35$ GHz with $1024$GB RAM.
Experiments on set B are conducted on machine B with $k \in \{2,8,16,64\}$, $\varepsilon = 0.03$, five seeds and a time limit of two hours.
Here, we use the full 64 cores.

Each partitioner optimizes the connectivity metric, which we also refer to as the quality of a partition.
For each instance (hypergraph and number of blocks $k$), we aggregate running times using the arithmetic mean over all seeds.
To further aggregate over multiple instances, we use the harmonic mean for relative speedups, and the geometric mean for absolute running times.
Runs with imbalanced partitions are not excluded from aggregated running times.
For runs that exceeded the time limit, we use the time limit itself as running time in the aggregates.
In plots, we mark these instances with \ClockLogo~if \emph{all} runs of that algorithm timed out, and similarly use \ding{55} to mark instances for which all runs produced imbalanced partitions.

To compare the solution quality of different algorithms, we use \emph{performance profiles}~\cite{PERFORMANCE-PROFILES}.
Let $\mathcal{A}$ be the set of all algorithms we want to compare, $\mathcal{I}$ the set of instances, and $q_{A}(I)$ the quality of algorithm
$A \in \mathcal{A}$ on instance $I \in \mathcal{I}$.
For each algorithm $A$, we plot the fraction of instances ($y$-axis) for which $q_A(I) \leq \tau \cdot \min_{A' \in \mathcal{A}}q_{A'}(I)$, where $\tau$ is on the $x$-axis.
Achieving higher fractions at lower $\tau$-values is considered better.
For $\tau = 1$, the $y$-value indicates the percentage of instances for which an algorithm performs best.
Note that these plots relate the quality of an algorithm to the best solution and thus do not permit a full ranking of three or more algorithms.

We additionally perform Wilcoxon signed rank tests~\cite{WILCOXON} to determine whether or not the differences between partitioners with similar quality are statistically significant.
At a $1\%$ significance level ($p \le 0.01$), a $Z$-score with $|Z| > 2.576$ is considered significant~\cite[p.~180]{WilcoxonZValues}.
Higher $p$-values also allow higher $Z$-score thresholds.

\subsection{Maximum Batch Size}

\begin{figure*}
  \centering
  \hspace{-1cm}
	\begin{minipage}{.32\textwidth}
		\centering
		\externalizedfigure{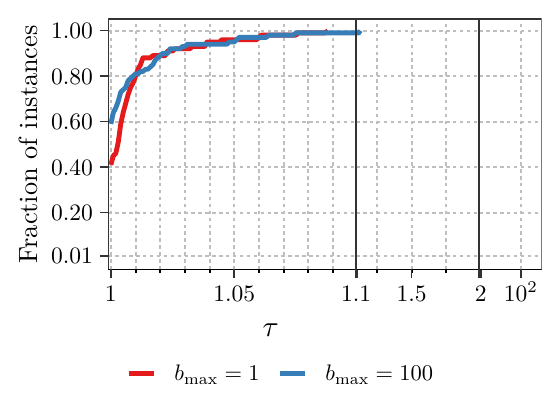}{experiments/maximum_batch_size/bmax_1_vs_100.tex} %
	\end{minipage} %
  \hspace{0.3cm}
	\begin{minipage}{.32\textwidth}
		\centering
		\externalizedfigure{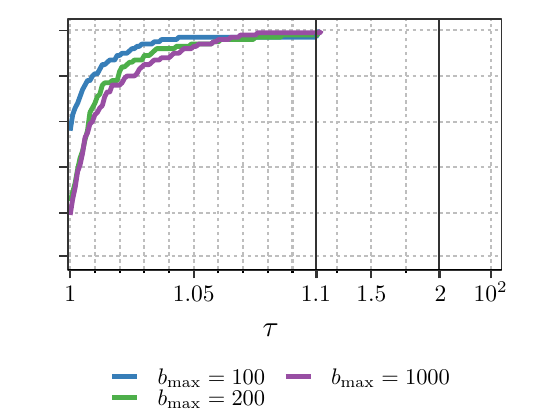}{experiments/maximum_batch_size/bmax_100_vs_200_vs_1000.tex} %
	\end{minipage} %
	\begin{minipage}{.32\textwidth}
		\centering
		\externalizedfigure{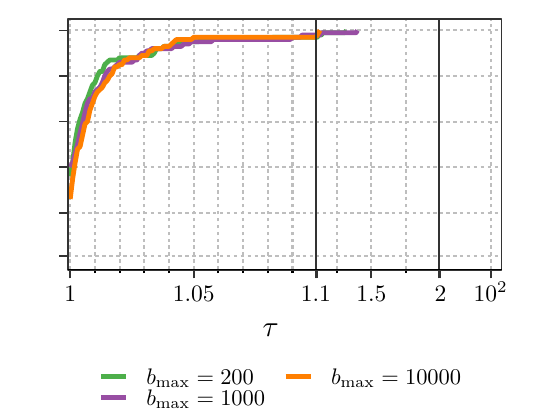}{experiments/maximum_batch_size/bmax_200_vs_1000_vs_10000.tex} %
	\end{minipage} %
	\vspace{-0.75cm}
  \caption{Performance profiles comparing the solution quality of \Partitioner{\mtkahyparnew} with different batch size
    values $\maxbatchsize$ on the $25$ instance subset of set B.}
  \label{fig:maximum_batch_size}
\end{figure*}

In Figure~\ref{fig:maximum_batch_size}, we compare the impact of different batch size values
$\splitatcommas{\maxbatchsize \in \{1,100,200,1000,10000\}}$ on the the solution quality on a subset of set B composed
of the $10$ smallest sparse matrices and $15$ smallest SAT instaces.
For this experiment, the minimum number of boundary vertices $\beta$ is set to $0$, so that refinement
is performed after each batch, and thus scalability depends only on $\maxbatchsize$.
Each run uses 20 threads.

We see that the different values yield roughly the same performance.
Comparing $\maxbatchsize = 1$ and $\maxbatchsize = 100$ shows a small advantage for $\maxbatchsize = 100$, which is not deemed significant with $Z = 0.92373$ and $p = 0.3556$.
Using $\maxbatchsize = 100$ is slightly better than $\maxbatchsize = 200$, but again the differences are not statistically significant ($Z = -2.1861$ and $p = 0.02881$).
Finally, there is no visible difference between $\maxbatchsize = 200, \maxbatchsize = 1000$ and $\maxbatchsize = 10000$, which is supported by the significance test for $\maxbatchsize = 200$ vs $\maxbatchsize = 1000$ with $Z = -0.60722$ and $p = 0.5437$.

In Table~\ref{tbl:gmean_running_times}, we report geometric mean running times of the different configurations.
The fastest configuration is $\maxbatchsize = 1000$, which is $1.64$ times faster than $\maxbatchsize = 100$ and $13.06$
times faster than $\maxbatchsize = 1$.
We therefore choose $\maxbatchsize = 1000$ for all further experiments.

\subsection{Scalability}

\begin{figure*}
	\centering
	\vspace{-1.65cm}
	\externalizedfigure{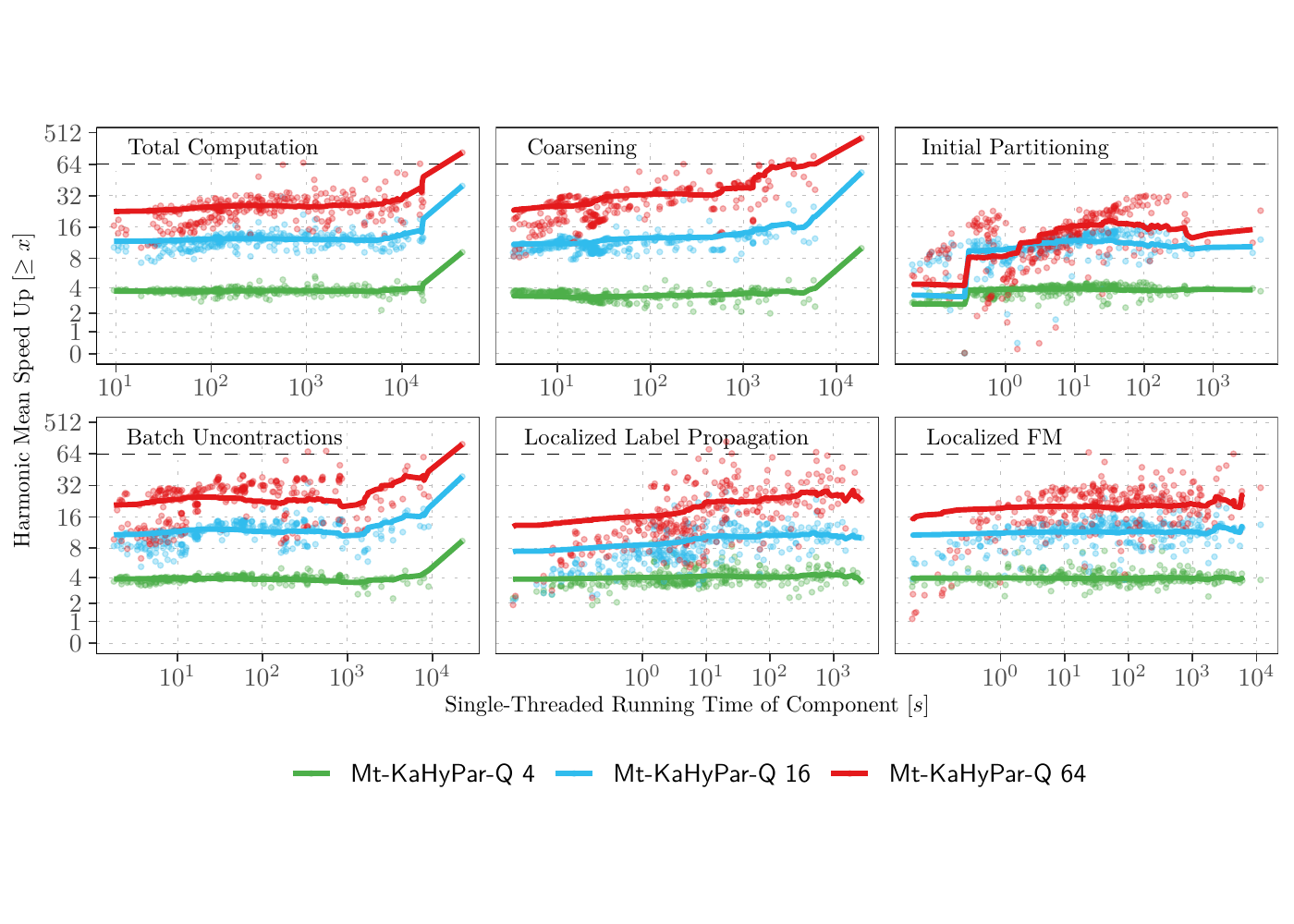}{experiments/scalability_experiments/speedup.tex} %
	\vspace{-2cm}
	\caption{Self-relative speedups for the different components of \Partitioner{\mtkahyparnew} on the 77 instance subset of set B. Harmonic mean speedups for instances with sequential running time $\geq x$.}
	\label{fig:speedup}
\end{figure*}

In Figure~\ref{fig:speedup} and Table~\ref{tbl:hmean_speed_ups}, we summarize self-relative speedups for several algorithmic components of \SeqAlgo{\mtkahyparnew} with
varying number of threads $p \in \{4,16,64\}$, on a subset of set B ($77$ out of $94$ hypergraphs\footnote{
The subset contains all hypergraphs on which \ExpAlgo{\mtkahyparnew}{64} was able to complete in under $800$ seconds
for all $k \in \{2,8,16,64\}$. This experiment still took $6$ weeks on machine B.}).
In the plot, we represent the speedup of each instance as a point and the cumulative harmonic mean speedup over
all instances with a single-threaded running time $\ge x$ seconds with a line.

\subparagraph*{Speedups.}
The overall harmonic mean speedup of \SeqAlgo{\mtkahyparnew} is $\placeholder{speedup}{speedUpTotalTimeP4S0}$
for $t = 4$, $\placeholder{speedup}{speedUpTotalTimeP16S0}$ for $t = 16$ and $\placeholder{speedup}{speedUpTotalTimeP64S0}$
for $t = 64$. If we only consider instances with a single-threaded running time $\ge 100$s, we achieve a harmonic mean
speed up of $\placeholder{speedup}{speedUpTotalTimeP64S100}$ for $t = 64$.

The scaling behavior of coarsening, batch uncontraction and localized FM are crucial to the overall
scalability of our framework, since they account for more than 60\% of the total
partitioning time on $\placeholder{share}{numCombinedInstancesWithShareGreater60P1}$\% of the instances for $t = 1$.
Coarsening and batch uncontractions both have similar speedups with $\placeholder{speedup}{speedUpCoarseningP16S100}$
and $\placeholder{speedup}{speedUpBatchUncontractionP16S100}$ for $t = 16$, respectively, whereas coarsening performs
significantly better with $t = 64$ ($\placeholder{speedup}{speedUpCoarseningP64S100}$ vs
$\placeholder{speedup}{speedUpBatchUncontractionP64S100}$).

Both localized refinement algorithms yield reliable speedups for an increasing number of threads.
However, since we run label propagation before FM, it scales slightly better for $t = 64$, as it could potentially remove boundary vertices from the cut, which then reduces the available parallelism for FM.
Initial partitioning shows the least promising speedups of all components, but is substantially faster, running in less than $100$ seconds on $\placeholder{speedup}{percentageInstancesSmallerInitialPartitioningS100}$\% of the instances for $t = 1$ (compare with last row of Table~\ref{tbl:hmean_speed_ups}).

On some instances we obtain super-linear speedups.
This is not surprising since the algorithms are non-deterministic.
Among other things, we have observed refinement converging in fewer rounds.
The extreme outliers, however, are caused by coarsening and batch uncontractions (up to 412), where non-deterministic contraction decisions cause highly varying running times (for the same number of threads).
Here, coarsening slows down once only few vertices are left, so that adding more cores could benefit from additional small non-shared caches.

\begin{table}
	\centering
	\caption{
		Harmonic mean speedups for different components: total computation (T), coarsening (C), initial partitioning (IP), uncontractions (BU), label propagation (LP) and FM.
		The last row shows how many instances took longer than 100 seconds in a certain phase using one thread.
		The middle row shows speedups for just these instances.
	}
	\label{tbl:hmean_speed_ups}
	\begin{tabular}{lr|rrrrrr}
		\multicolumn{2}{r|}{Num. Threads} & T & C & IP & BU & LP & FM \\\hline
		\multirow{3}{*}{All}        & $4$ & $\placeholder{speedup}{speedUpTotalTimeP4S0}$ & $\placeholder{speedup}{speedUpCoarseningP4S0}$ &
		$\placeholder{speedup}{speedUpInitialPartitiongP4S0}$ & $\placeholder{speedup}{speedUpBatchUncontractionP4S0}$ &
		$\placeholder{speedup}{speedUpLabelPropagationP4S0}$ & $\placeholder{speedup}{speedUpFMP4S0}$ \\
		& $16$ & $\placeholder{speedup}{speedUpTotalTimeP16S0}$ & $\placeholder{speedup}{speedUpCoarseningP16S0}$ &
		$\placeholder{speedup}{speedUpInitialPartitiongP16S0}$ & $\placeholder{speedup}{speedUpBatchUncontractionP16S0}$ &
		$\placeholder{speedup}{speedUpLabelPropagationP16S0}$ & $\placeholder{speedup}{speedUpFMP16S0}$  \\
		& $64$ & $\placeholder{speedup}{speedUpTotalTimeP64S0}$ & $\placeholder{speedup}{speedUpCoarseningP64S0}$ &
		$\placeholder{speedup}{speedUpInitialPartitiongP64S0}$ & $\placeholder{speedup}{speedUpBatchUncontractionP64S0}$ &
		$\placeholder{speedup}{speedUpLabelPropagationP64S0}$ & $\placeholder{speedup}{speedUpFMP64S0}$  \\\hline
		\multirow{3}{*}{$\ge 100$s} & $4$  & $\placeholder{speedup}{speedUpTotalTimeP4S100}$ & $\placeholder{speedup}{speedUpCoarseningP4S100}$ &
		$\placeholder{speedup}{speedUpInitialPartitiongP4S100}$ & $\placeholder{speedup}{speedUpBatchUncontractionP4S100}$ &
		$\placeholder{speedup}{speedUpLabelPropagationP4S100}$ & $\placeholder{speedup}{speedUpFMP4S100}$  \\
		& $16$ & $\placeholder{speedup}{speedUpTotalTimeP16S100}$ & $\placeholder{speedup}{speedUpCoarseningP16S100}$ &
		$\placeholder{speedup}{speedUpLabelPropagationP16S100}$ & $\placeholder{speedup}{speedUpFMP16S100}$ &
		$\placeholder{speedup}{speedUpInitialPartitiongP16S100}$ & $\placeholder{speedup}{speedUpBatchUncontractionP16S100}$ \\
		& $64$ & $\placeholder{speedup}{speedUpTotalTimeP64S100}$ & $\placeholder{speedup}{speedUpCoarseningP64S100}$ &
		$\placeholder{speedup}{speedUpInitialPartitiongP64S100}$ & $\placeholder{speedup}{speedUpBatchUncontractionP64S100}$ &
		$\placeholder{speedup}{speedUpLabelPropagationP64S100}$ & $\placeholder{speedup}{speedUpFMP64S100}$  \\\hline
		\multicolumn{2}{r|}{Inst. $\ge 100$s $[\%]$} & $\placeholder{speedup}{percentageInstancesGreaterTotalTimeS100}$ & $\placeholder{speedup}{percentageInstancesGreaterCoarseningS100}$ &
		$\placeholder{speedup}{percentageInstancesGreaterInitialPartitioningS100}$ & $\placeholder{speedup}{percentageInstancesGreaterBatchUncontractionS100}$ &
		$\placeholder{speedup}{percentageInstancesGreaterLabelPropagationS100}$ & $\placeholder{speedup}{percentageInstancesGreaterFMS100}$  \\
	\end{tabular}
	\vspace*{-0.4cm}
\end{table}

\subparagraph*{Quality with Increasing Number of Threads.}

Figure~\ref{fig:quality_comparison_set_b} (right), shows that increasing the number of threads does adversely affect the solution quality of \Partitioner{\mtkahyparnew}, but only by a very small amount.
We observed that the localized searches interfere with one another, which may reasonably explain this behavior, and also explain why this issue does not arise for \Partitioner{\mtkahyparold}~\cite{MT-KAHYPAR}.
If the number of seed vertices of FM is sufficiently high, the searches operate in different areas of the hypergraph to a certain extent, since the seeds are randomly shuffled.
A possible remedy would be to further diversify the uncontraction batches.

\subsection{Comparison with other Algorithms.}

\begin{figure*}
  \centering
  \hspace{-1cm}
	\begin{minipage}{.32\textwidth}
		\centering
		\externalizedfigure{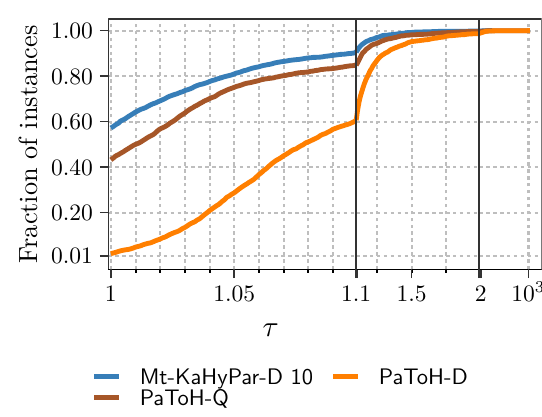}{experiments/kahypar_benchmark_set/quality_plot_patoh_vs_mt_kahypar_fast.tex} %
	\end{minipage} %
  \hspace{0.3cm}
	\begin{minipage}{.32\textwidth}
		\centering
		\externalizedfigure{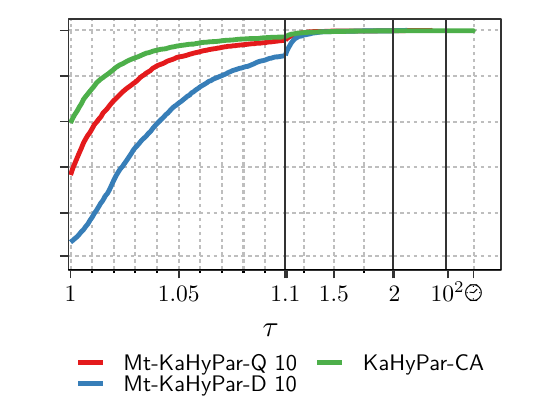}{experiments/kahypar_benchmark_set/quality_plot_kahypar_ca_vs_mt_kahypar_strong.tex} %
	\end{minipage} %
	\begin{minipage}{.32\textwidth}
		\centering
		\externalizedfigure{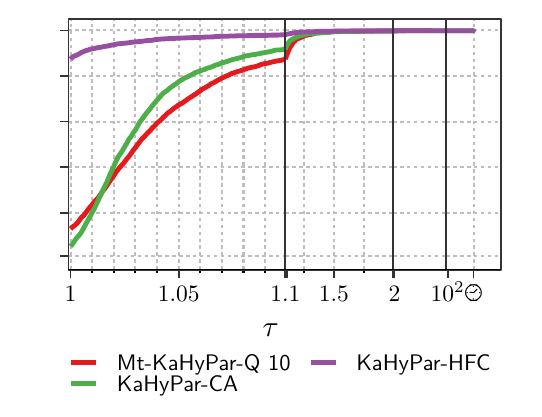}{experiments/kahypar_benchmark_set/quality_plot_kahypar_hfc_vs_mt_kahypar_strong.tex} %
	\end{minipage} %
	\vspace{-0.75cm}
	\caption{Performance profiles comparing the solution quality of \Partitioner{\mtkahyparold} (left and middle) and \Partitioner{\mtkahyparnew} (middle and right)
		with different sequential partitioners on set A.}
	\label{fig:quality_comparison_set_a}
\end{figure*}

\begin{figure*}
	\centering
  \hspace{-1cm}
	\begin{minipage}{.32\textwidth}
		\centering
		\externalizedfigure{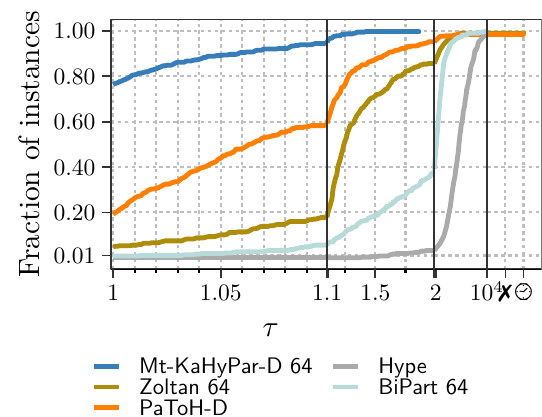}{experiments/big_benchmark_set/quality_plot_all_vs_mt_kahypar_fast.tex} %
	\end{minipage} %
  \hspace{0.3cm}
	\begin{minipage}{.32\textwidth}
		\centering
		\externalizedfigure{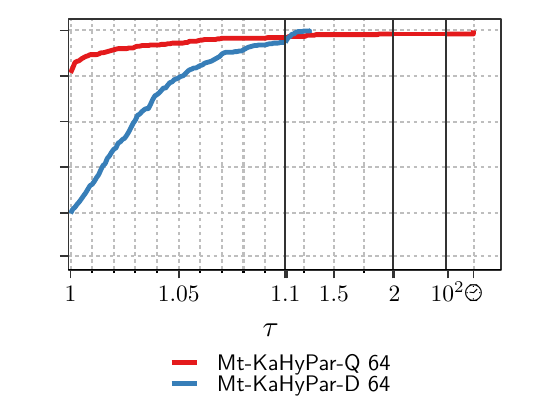}{experiments/big_benchmark_set/quality_plot_mt_kahypar_strong_vs_fast.tex} %
	\end{minipage} %
	\begin{minipage}{.32\textwidth}
		\centering
		\externalizedfigure{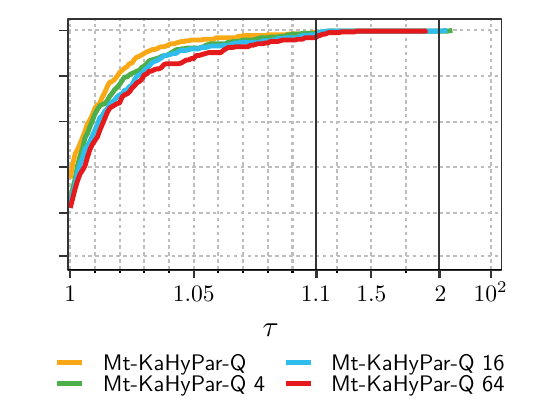}{experiments/scalability_experiments/quality_plot_big_scaling.tex} %
	\end{minipage} %
	\vspace{-0.55cm}
	\caption{Performance profiles comparing the solution quality of \Partitioner{\mtkahyparold} (left) and \Partitioner{\mtkahyparnew} (middle) with
		different parallel and fast sequential partitioners on set B, as well as \Partitioner{\mtkahyparnew} with varying number of threads (right).}
	\label{fig:quality_comparison_set_b}
\end{figure*}

Figures~\ref{fig:quality_comparison_set_a} and~\ref{fig:quality_comparison_set_b} compare the solution
quality of \Partitioner{\mtkahyparold} and \Partitioner{\mtkahyparnew} with
different partitioners.

\subparagraph*{Quality Comparison on Set A.}

We first look at set $A$.
\Partitioner{\mtkahyparold} finds better partitions than either \Partitioner{PaToH} variant, while \Partitioner{\mtkahyparnew} finds better partitions than \Partitioner{\mtkahyparold}, and partitions of similar quality as \Partitioner{KaHyPar-CA}.
\Partitioner{KaHyPar-HFC} still computes the highest quality partitions, due to the additional flow-based refinement~\cite{KAHYPAR-HFC,KAHYPAR-MF}.
In an individual comparison, \ExpAlgo{\mtkahyparnew}{10} finds better partitions than \Partitioner{PaToH-D}, \ExpAlgo{\mtkahyparold}{10}, \Partitioner{PaToH-Q},
\Partitioner{KaHyPar-CA}, \Partitioner{KaHyPar-HFC} on
$\placeholder{sequential}{mtKaHyParStrongvsPaToHDMtKaHyParStrong10Tau100}\%$,
$\placeholder{sequential}{mtKaHyParStrongvsmtKaHyParFastMtKaHyParStrong10Tau100}\%$,
$\placeholder{sequential}{mtKaHyParStrongvsPaToHQMtKaHyParStrong10Tau100}\%$,
$\placeholder{sequential}{mtKaHyParStrongvsKaHyParCAMtKaHyParStrong10Tau100}\%$ and
$\placeholder{sequential}{mtKaHyParStrongvsKaHyParHFCMtKaHyParStrong10Tau100}\%$ of
the instances, respectively.
The Wilcoxon test indicates that the difference between \ExpAlgo{\mtkahyparnew}{10} and \Partitioner{KaHyPar-CA} is not statistically significant ($Z = 1.1702, p = 0.2419$).
\ExpAlgo{\mtkahyparold}{10}
performs significantly better than \Partitioner{PaToH-Q} ($Z = -16.478, p < 2.2e-16$) and is also faster
than \Partitioner{PaToH-D} in the geometric mean ($0.96$s vs $1.17$s, see Table~\ref{tbl:gmean_running_times}).

\subparagraph*{Quality Comparison on Set B}

We now turn to the evaluation on set B.
\ExpAlgo{\mtkahyparold}{64} produces partitions with better quality than \SeqAlgo{Hype}, \ExpAlgo{BiPart}{64}, \ExpAlgo{Zoltan}{64}, \SeqAlgo{PaToH-D} on
$\placeholder{parallel}{mtKaHyParFastvsHypeMtKaHyParFast64Tau100}\%$,
$\placeholder{parallel}{mtKaHyParFastvsBiPartMtKaHyParFast64Tau100}\%$,
$\placeholder{parallel}{mtKaHyParFastvsZoltanMtKaHyParFast64Tau100}\%$ and
$\placeholder{parallel}{mtKaHyParFastvsPaToHDMtKaHyParFast64Tau100}\%$ of the instances, respectively.
Moreover, \ExpAlgo{\mtkahyparnew}{64} outperforms \ExpAlgo{\mtkahyparold}{64} in terms of solutions quality on
$\placeholder{parallel}{mtKaHyParFastvsMtKahyparStrongMtKaHyParStrong64Tau100}\%$ of the instances.

\subparagraph*{Running Times.}

\begin{figure*}
	\centering
	\begin{minipage}{.49\textwidth}
		\centering
		\externalizedfigure{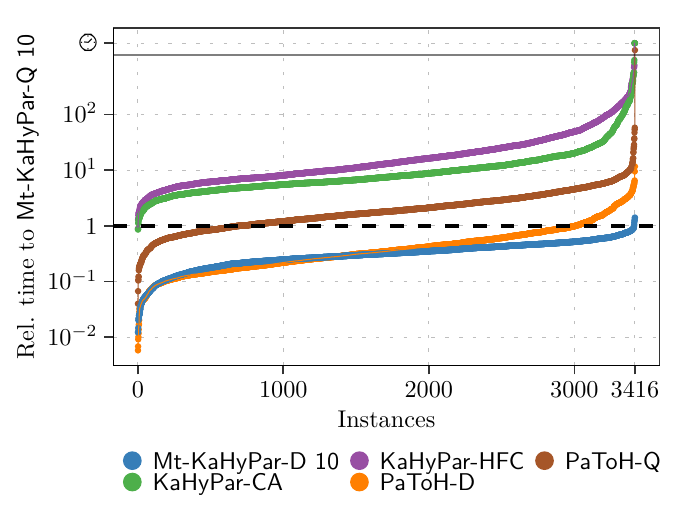}{experiments/kahypar_benchmark_set/kahypar_relative_runtime.tex} %
	\end{minipage} %
	\begin{minipage}{.49\textwidth}
		\centering
		\externalizedfigure{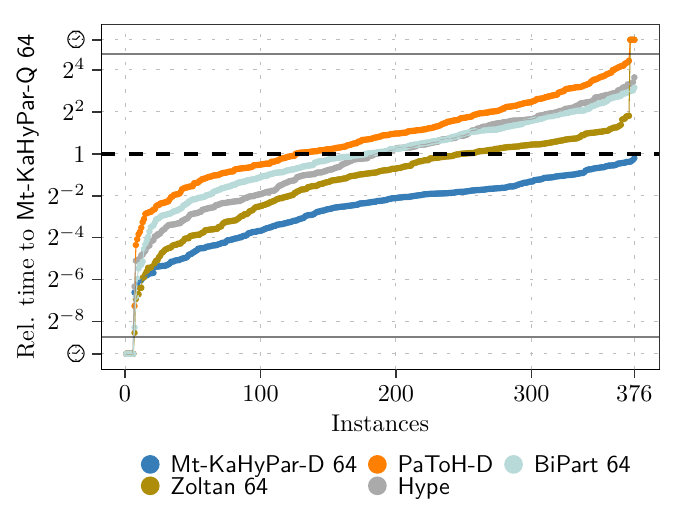}{experiments/big_benchmark_set/big_relative_runtime.tex} %
	\end{minipage} %
	\vspace{-0.6cm}
	\caption{Running times relative to \ExpAlgo{\mtkahyparnew}{10} on set A
		(left), respectively~\ExpAlgo{\mtkahyparnew}{64} on set B (right). The \ClockLogo~axis markers represent timeouts for the baseline \Partitioner{\mtkahyparnew} (at the bottom) or the compared algorithm (at the top).}
	\label{fig:relative_runtime_comparison}
\end{figure*}

Figure~\ref{fig:relative_runtime_comparison} shows running times relative to \Partitioner{\mtkahyparnew}.
On set A, \Partitioner{\mtkahyparnew} is faster than \Partitioner{KaHyPar-HFC}, \Partitioner{KaHyPar-CA} and \Partitioner{PaToH-Q}, but slower than \Partitioner{\mtkahyparold} and \Partitioner{PaToH-D}.
On set B, \Partitioner{\mtkahyparold} is substantially faster than \Partitioner{\mtkahyparnew}, whereas \Partitioner{Hype}, \Partitioner{BiPart} and \Partitioner{Zoltan} are only moderately faster, while \Partitioner{PaToH-D} is slower.
In Table~\ref{tbl:gmean_running_times}, we additionally report geometric mean running times.

While developing \Partitioner{\mtkahyparnew}, we made numerous improvements on the implementation level that also affect \Partitioner{\mtkahyparold}.
On set A, the new version of \Partitioner{\mtkahyparold} is a factor of $1.57$ faster ($1.5$s vs $0.95$s geometric mean) than the old version~\cite{MT-KAHYPAR}, and a factor of $1.26$ faster on set B ($6.14$s vs $4.89$s).

In conclusion, \Partitioner{\mtkahyparold} is the method of choice for partitioning hypergraphs very quickly with good quality.
For high quality partitions, \Partitioner{KaHyPar-HFC} remains the best choice on medium-size instances ($\leq 10$ million pins), whereas \Partitioner{\mtkahyparnew} should be used on large hypergraphs, where the running time of a sequential algorithm becomes prohibitive.
We plan to close this gap by parallelizing the flow-based refinement of \Partitioner{KaHyPar-HFC}.

\begin{table}
	\caption{Geometric mean running times.}
	\label{tbl:gmean_running_times}
	\centering
    \vspace{-0.3cm}
	\scalebox{0.9}{
		\begin{tabular}{lr lr  lr}
			\multicolumn{2}{c}{Set A} & \multicolumn{2}{c}{Set B} & \multicolumn{2}{c}{Subset of Set B}  \\
			Partitioner & $t[s]$ & Partitioner & $t[s]$  & $\maxbatchsize$ & $t[s]$  \\
			\cmidrule(lr){1-2} \cmidrule(lr){3-4} \cmidrule(lr){5-6}
			\ExpAlgo{\mtkahyparold}{10}     & $\placeholder{sequential}{gmeanTimeMtKaHyParFast10}$     & \ExpAlgo{\mtkahyparold}{64}   & $\placeholder{parallel}{gmeanTimeMtKaHyParFast64}$   & $1$     & $\placeholder{batch_size}{gmeanbmax1}$ \\
			\Partitioner{PaToH-D}           & $\placeholder{sequential}{gmeanTimePaToHD}$              & \ExpAlgo{Zoltan}{64}          & $\placeholder{parallel}{gmeanTimeZoltan64}$          & $100$   & $\placeholder{batch_size}{gmeanbmax100}$ \\
			\ExpAlgo{\mtkahyparnew}{10}     & $\placeholder{sequential}{gmeanTimeMtKaHyParStrong10}$   & \Partitioner{Hype}            & $\placeholder{parallel}{gmeanTimeHype}$              & $200$   & $\placeholder{batch_size}{gmeanbmax200}$ \\
			\Partitioner{PaToH-Q}           & $\placeholder{sequential}{gmeanTimePaToHQ}$              & \ExpAlgo{BiPart}{64}         & $\placeholder{parallel}{gmeanTimeBiPart64}$           & $1000$  & $\placeholder{batch_size}{gmeanbmax1000}$ \\
			\Partitioner{KaHyPar-CA}        & $\placeholder{sequential}{gmeanTimeKaHyParCA}$           & \ExpAlgo{\mtkahyparnew}{64}   & $\placeholder{parallel}{gmeanTimeMtKaHyParStrong64}$ & $10000$ & $\placeholder{batch_size}{gmeanbmax10000}$ \\
			\Partitioner{KaHyPar-HFC}       & $\placeholder{sequential}{gmeanTimeKaHyParHFC}$          & \Partitioner{PaToH-D}         & $\placeholder{parallel}{gmeanTimePaToHD}$            & \\
    \end{tabular}
  }
\vspace{-0.5cm}
\end{table}

\subparagraph*{Comparison with SHP.}

Since we could not include \Partitioner{SHP} in the regular comparison, we ran \Partitioner{Mt-KaHyPar} on the available real-world instances and compare our results to the results reported in the publication~\cite{SHP}.
\Partitioner{SHP} optimizes the \emph{fanout} objective, which is equivalent to connectivity on unweighted instances, since $(\lambda-1)(\Partition) = (\text{fanout}(\Partition)\cdot |E|) - |E|$.
In Table~\ref{tbl:social_hash_comparison}, we report connectivity values (converted from fanout for \Partitioner{SHP}) and running times for \Partitioner{SHP-k} (direct k-way), \Partitioner{SHP-2} (recursive bipartitioning), \Partitioner{\mtkahyparold} and \Partitioner{\mtkahyparnew}, all using 64 threads, with $\epsilon = 0.05$ and $k \in \{2, 8, 32, 128\}$.
The instances are ordered from smallest to largest.%
\footnote{Detailed instance properties are available at \url{http://algo2.iti.kit.edu/heuer/nlevel/}.}
On most of the instances, \Partitioner{Mt-KaHyPar} computes substantially better partitions.
The difference is most pronounced for the web instances.
Only for the two smallest instances email--Enron with $k=2$, and soc--Epinions for $k \geq 8$ \Partitioner{SHP} computes slightly better partitions.
Since the authors of \Partitioner{SHP} use different machines (4 x 16 threads), which are older than ours, and only report times for the 2 largest instances and $k = 32$, the comparison of running times is limited.
Even though \Partitioner{SHP} is a flat algorithm (no multilevel) and performs only label propagation, \Partitioner{\mtkahyparold} is faster in the reported numbers.
Since their machine is older, we estimate that the running times of \Partitioner{SHP} and \Partitioner{\mtkahyparold} are relatively similar, whereas \Partitioner{\mtkahyparnew} is about a factor of 5--10 slower.

\begin{table}
  \caption{Comparison with \Partitioner{SHP}. Best solutions in bold.}\label{tbl:social_hash_comparison}
  \centering
	\vspace{-0.25cm}
	\scalebox{0.75}{
		\begin{tabular}{lr rrrr r}
			&             & \multicolumn{4}{ c }{$\lambda - 1$} & \multicolumn{1}{c}{$t[s]$} \\
			\cmidrule(lr){3-6} \cmidrule(lr){7-7}
			$H$ & Partitioner & $k = 2$ & $k = 8$ & $k = 32$ & $k = 128$ & $k = 32$ \\
			\midrule
			\parbox[t]{2mm}{\multirow{4}{*}{\rotatebox[origin=c]{90}{\footnotesize email--Enron}}}
			& \Partitioner{SHP-2}             & $\pmb{\placeholder{shp}{SHP2emailEnronK2}}$    &
			$\placeholder{shp}{SHP2emailEnronK8}$    &
			$\placeholder{shp}{SHP2emailEnronK32}$  &
			$\placeholder{shp}{SHP2emailEnronK128}$ &
			$\placeholder{shp}{SHP2emailEnronK32Time}$ \\
			& \Partitioner{SHP-K}             & $\placeholder{shp}{SHPKemailEnronK2}$    &
			$\placeholder{shp}{SHPKemailEnronK8}$    &
			$\placeholder{shp}{SHPKemailEnronK32}$  &
			$\placeholder{shp}{SHPKemailEnronK128}$ &
			$\placeholder{shp}{SHP2emailEnronK32Time}$  \\
			& \Partitioner{\mtkahyparold}   & $\placeholder{shp}{MtKaHyParFastemailEnronK2}$   &
			$\placeholder{shp}{MtKaHyParFastemailEnronK8}$   &
			$\placeholder{shp}{MtKaHyParFastemailEnronK32}$  &
			$\placeholder{shp}{MtKaHyParFastemailEnronK128}$ &
			$\placeholder{shp}{MtKaHyParFastemailEnronK32Time}$  \\
			& \Partitioner{\mtkahyparnew} & $\placeholder{shp}{MtKaHyParStrongemailEnronK2}$   &
			$\pmb{\placeholder{shp}{MtKaHyParStrongemailEnronK8}}$   &
			$\pmb{\placeholder{shp}{MtKaHyParStrongemailEnronK32}}$  &
			$\pmb{\placeholder{shp}{MtKaHyParStrongemailEnronK128}}$ &
			$\placeholder{shp}{MtKaHyParStrongemailEnronK32Time}$ \\
			\midrule
			\parbox[t]{2mm}{\multirow{4}{*}{\rotatebox[origin=c]{90}{\footnotesize soc--Epinions}}}
			& \Partitioner{SHP-2}             & $\placeholder{shp}{SHP2socEpinionsK2}$   &
			$\placeholder{shp}{SHP2socEpinionsK8}$   &
			$\placeholder{shp}{SHP2socEpinionsK32}$  &
			$\placeholder{shp}{SHP2socEpinionsK128}$  &
			$\placeholder{shp}{SHP2socEpinionsK32Time}$ \\
			& \Partitioner{SHP-K}             & $\placeholder{shp}{SHPKsocEpinionsK2}$   &
			$\pmb{\placeholder{shp}{SHPKsocEpinionsK8}}$   &
			$\pmb{\placeholder{shp}{SHPKsocEpinionsK32}}$  &
			$\pmb{\placeholder{shp}{SHPKsocEpinionsK128}}$ &
			$\placeholder{shp}{SHPKsocEpinionsK32Time}$ \\
			& \Partitioner{\mtkahyparold}   & $\placeholder{shp}{MtKaHyParFastsocEpinionsK2}$   &
			$\placeholder{shp}{MtKaHyParFastsocEpinionsK8}$   &
			$\placeholder{shp}{MtKaHyParFastsocEpinionsK32}$  &
			$\placeholder{shp}{MtKaHyParFastsocEpinionsK128}$ &
			$\placeholder{shp}{MtKaHyParFastsocEpinionsK32Time}$ \\
			& \Partitioner{\mtkahyparnew} & $\pmb{\placeholder{shp}{MtKaHyParStrongsocEpinionsK2}}$   &
			$\placeholder{shp}{MtKaHyParStrongsocEpinionsK8}$   &
			$\placeholder{shp}{MtKaHyParStrongsocEpinionsK32}$  &
			$\placeholder{shp}{MtKaHyParStrongsocEpinionsK128}$ &
			$\placeholder{shp}{MtKaHyParStrongsocEpinionsK32Time}$  \\
			\midrule
			\parbox[t]{2mm}{\multirow{4}{*}{\rotatebox[origin=c]{90}{\footnotesize web--Stanford}}}
			& \Partitioner{SHP-2}             & $\placeholder{shp}{SHP2webStanfordK2}$   &
			$\placeholder{shp}{SHP2webStanfordK8}$   &
			$\placeholder{shp}{SHP2webStanfordK32}$  &
			$\placeholder{shp}{SHP2webStanfordK128}$ &
			$\placeholder{shp}{SHP2webStanfordK32Time}$  \\
			& \Partitioner{SHP-K}             & $\placeholder{shp}{SHPKwebStanfordK2}$   &
			$\placeholder{shp}{SHPKwebStanfordK8}$   &
			$\placeholder{shp}{SHPKwebStanfordK32}$  &
			$\placeholder{shp}{SHPKwebStanfordK128}$ &
			$\placeholder{shp}{SHPKwebStanfordK32Time}$ \\
			& \Partitioner{\mtkahyparold}   & $\pmb{\placeholder{shp}{MtKaHyParFastwebStanfordK2}}$   &
			$\placeholder{shp}{MtKaHyParFastwebStanfordK8}$   &
			$\placeholder{shp}{MtKaHyParFastwebStanfordK32}$  &
			$\placeholder{shp}{MtKaHyParFastwebStanfordK128}$  &
			$\placeholder{shp}{MtKaHyParFastwebStanfordK32Time}$ \\
			& \Partitioner{\mtkahyparnew} & $\placeholder{shp}{MtKaHyParStrongwebStanfordK2}$   &
			$\pmb{\placeholder{shp}{MtKaHyParStrongwebStanfordK8}}$   &
			$\pmb{\placeholder{shp}{MtKaHyParStrongwebStanfordK32}}$  &
			$\pmb{\placeholder{shp}{MtKaHyParStrongwebStanfordK128}}$ &
			$\placeholder{shp}{MtKaHyParStrongwebStanfordK32Time}$ \\
			\midrule
			\parbox[t]{2mm}{\multirow{4}{*}{\rotatebox[origin=c]{90}{\footnotesize web--BerkStan}}}
			& \Partitioner{SHP-2}             & $\placeholder{shp}{SHP2webBerkStanK2}$   &
			$\placeholder{shp}{SHP2webBerkStanK8}$   &
			$\placeholder{shp}{SHP2webBerkStanK32}$  &
			$\placeholder{shp}{SHP2webBerkStanK128}$ &
			$\placeholder{shp}{SHP2webBerkStanK32Time}$ \\
			& \Partitioner{SHP-K}             & $\placeholder{shp}{SHPKwebBerkStanK2}$   &
			$\placeholder{shp}{SHPKwebBerkStanK8}$   &
			$\placeholder{shp}{SHPKwebBerkStanK32}$  &
			$\placeholder{shp}{SHPKwebBerkStanK128}$ &
			$\placeholder{shp}{SHPKwebBerkStanK32Time}$ \\
			& \Partitioner{\mtkahyparold}   & $\pmb{\placeholder{shp}{MtKaHyParFastwebBerkStanK2}}$   &
			$\placeholder{shp}{MtKaHyParFastwebBerkStanK8}$   &
			$\placeholder{shp}{MtKaHyParFastwebBerkStanK32}$  &
			$\placeholder{shp}{MtKaHyParFastwebBerkStanK128}$ &
			$\placeholder{shp}{MtKaHyParFastwebBerkStanK32Time}$ \\
			& \Partitioner{\mtkahyparnew} & $\placeholder{shp}{MtKaHyParStrongwebBerkStanK2}$   &
			$\pmb{\placeholder{shp}{MtKaHyParStrongwebBerkStanK8}}$   &
			$\pmb{\placeholder{shp}{MtKaHyParStrongwebBerkStanK32}}$  &
			$\pmb{\placeholder{shp}{MtKaHyParStrongwebBerkStanK128}}$ &
			$\placeholder{shp}{MtKaHyParStrongwebBerkStanK32Time}$ \\
			\midrule
			\parbox[t]{2mm}{\multirow{4}{*}{\rotatebox[origin=c]{90}{\footnotesize soc--Pokec}}}
			& \Partitioner{SHP-2}             & $\placeholder{shp}{SHP2socPokecK2}$   &
			$\placeholder{shp}{SHP2socPokecK8}$   &
			$\placeholder{shp}{SHP2socPokecK32}$  &
			$\placeholder{shp}{SHP2socPokecK128}$  &
			$\placeholder{shp}{SHP2socPokecK32Time}$ \\
			& \Partitioner{SHP-K}             & $\placeholder{shp}{SHPKsocPokecK2}$   &
			$\placeholder{shp}{SHPKsocPokecK8}$   &
			$\placeholder{shp}{SHPKsocPokecK32}$  &
			$\placeholder{shp}{SHPKsocPokecK128}$  &
			$\placeholder{shp}{SHPKsocPokecK32Time}$ \\
			& \Partitioner{\mtkahyparold}   & $\placeholder{shp}{MtKaHyParFastsocPokecK2}$   &
			$\pmb{\placeholder{shp}{MtKaHyParFastsocPokecK8}}$   &
			$\placeholder{shp}{MtKaHyParFastsocPokecK32}$  &
			$\placeholder{shp}{MtKaHyParFastsocPokecK128}$  &
			$\placeholder{shp}{MtKaHyParFastsocPokecK32Time}$ \\
			& \Partitioner{\mtkahyparnew} & $\pmb{\placeholder{shp}{MtKaHyParStrongsocPokecK2}}$   &
			$\placeholder{shp}{MtKaHyParStrongsocPokecK8}$   &
			$\pmb{\placeholder{shp}{MtKaHyParStrongsocPokecK32}}$  &
			$\pmb{\placeholder{shp}{MtKaHyParStrongsocPokecK128}}$  &
			$\placeholder{shp}{MtKaHyParStrongsocPokecK32Time}$ \\
			\midrule
			\parbox[t]{2mm}{\multirow{4}{*}{\rotatebox[origin=c]{90}{\footnotesize soc--LJ}}}
			& \Partitioner{SHP-2}             & $\placeholder{shp}{SHP2socLiveJournalK2}$   &
			$\placeholder{shp}{SHP2socLiveJournalK8}$   &
			$\placeholder{shp}{SHP2socLiveJournalK32}$  &
			$\placeholder{shp}{SHP2socLiveJournalK128}$ &
			$\placeholder{shp}{SHP2socLiveJournalK32Time}$ \\
			& \Partitioner{SHP-K}             & $\placeholder{shp}{SHPKsocLiveJournalK2}$   &
			$\placeholder{shp}{SHPKsocLiveJournalK8}$   &
			$\placeholder{shp}{SHPKsocLiveJournalK32}$  &
			$\placeholder{shp}{SHPKsocLiveJournalK128}$  &
			$\placeholder{shp}{SHPKsocLiveJournalK32Time}$ \\
			& \Partitioner{\mtkahyparold}   & $\placeholder{shp}{MtKaHyParFastsocLiveJournalK2}$   &
			$\placeholder{shp}{MtKaHyParFastsocLiveJournalK8}$   &
			$\placeholder{shp}{MtKaHyParFastsocLiveJournalK32}$  &
			$\placeholder{shp}{MtKaHyParFastsocLiveJournalK128}$  &
			$\placeholder{shp}{MtKaHyParFastsocLiveJournalK32Time}$ \\
			& \Partitioner{\mtkahyparnew} & $\pmb{\placeholder{shp}{MtKaHyParStrongsocLiveJournalK2}}$   &
			$\pmb{\placeholder{shp}{MtKaHyParStrongsocLiveJournalK8}}$   &
			$\pmb{\placeholder{shp}{MtKaHyParStrongsocLiveJournalK32}}$  &
			$\pmb{\placeholder{shp}{MtKaHyParStrongsocLiveJournalK128}}$   &
			$\placeholder{shp}{MtKaHyParStrongsocLiveJournalK32Time}$ \\
		\end{tabular}
	}
	  \vspace{-0.5cm}
\end{table}

\section{Conclusion and Future Work}\label{s:conclusion}

In this paper, we demonstrated that powerful $n$-level algorithms for hypergraph partitioning can be parallelized very efficiently on shared-memory architectures, without significant sacrifices in solution quality.
Our experiments show that multilevel algorithms achieve much higher quality than recently proposed flat algorithms, while being competitive in running time.
This is particularly relevant for applications such as storage sharding where improvements directly translate to savings in cost and running time.

Future work on the algorithmic side includes parallelizing flow-based refinement techniques~\cite{REBAHFC, KAHYPAR-HFC}, as well as techniques to construct batches that reduce interference between localized searches.
Even more scalable (but also more challenging) would be an asynchronous version of the uncoarsening phase without synchronization after each batch, where uncontractions and refinement happen concurrently.
Perhaps transactional memory or similar mechanisms might help to maintain consistent partition states.





\bibliography{mt_kahypar}

\begin{thebibliography}{10}

\bibitem{KAHYPAR-K}
Yaroslav Akhremtsev, Tobias Heuer, Peter Sanders, and Sebastian Schlag.
\newblock Engineering a direct \emph{k}-way hypergraph partitioning algorithm.
\newblock In {\em 19th Workshop on Algorithm Engineering \& Experiments
  (ALENEX)}, pages 28--42. SIAM, 01 2017.

\bibitem{MT-KAHIP}
Yaroslav Akhremtsev, Peter Sanders, and Christian Schulz.
\newblock High-quality shared-memory graph partitioning.
\newblock In {\em European Conference on Parallel Processing (Euro-Par)}, pages
  659--671. Springer, 8 2017.

\bibitem{ISPD98}
Charles~J. Alpert.
\newblock The ispd98 circuit benchmark suite.
\newblock In {\em International Symposium on Physical Design (ISPD)}, pages
  80--85, 4 1998.

\bibitem{ALPERT-SURVEY}
Charles~J. Alpert and Andrew~B. Kahng.
\newblock Recent directions in netlist partitioning: A survey.
\newblock {\em Integration}, 19(1-2):1--81, 1995.

\bibitem{kPaToH}
Cevdet Aykanat, Berkant~Barla Cambazoglu, and Bora U{\c{c}}ar.
\newblock Multi-level direct $k$-way hypergraph partitioning with multiple
  constraints and fixed vertices.
\newblock {\em Journal of Parallel and Distributed Computing}, 68(5):609--625,
  2008.
\newblock \href {https://doi.org/10.1016/j.jpdc.2007.09.006}
  {\path{doi:10.1016/j.jpdc.2007.09.006}}.

\bibitem{GRAPH-SURVEY}
David~A. Bader, Henning Meyerhenke, Peter Sanders, and Dorothea Wagner.
\newblock {\em Graph Partitioning and Graph Clustering}, volume 588.
\newblock American Mathematical Society Providence, RI, 2013.

\bibitem{SAT14}
Anton Belov, Daniel Diepold, Marijn Heule, and Matti J{\"{a}}rvisalo.
\newblock The sat competition 2014.
\newblock \url{http://www.satcompetition.org/2014/}, 2014.

\bibitem{BUI}
Thang~N. Bui and Curt Jones.
\newblock A heuristic for reducing fill-in in sparse matrix factorization.
\newblock In {\em 6th {SIAM} Conference on Parallel Processing for Scientific
  Computing (PPSC)}, pages 445--452, 1993.

\bibitem{WilcoxonZValues}
Michael~J. Campbell and Thomas~D.V. Swinscow.
\newblock {\em Statistics at Square One}.
\newblock BMJ Publishing Group, 2009.

\bibitem{PATOH}
Ümit~V. Catalyurek and Cevdet Aykanat.
\newblock Hypergraph-partitioning-based decomposition for parallel
  sparse-matrix vector multiplication.
\newblock {\em IEEE Transactions on Parallel and Distributed Systems},
  10(7):673--693, 1999.

\bibitem{schism}
Carlo Curino, Yang Zhang, Evan P.~C. Jones, and Samuel Madden.
\newblock Schism: a workload-driven approach to database replication and
  partitioning.
\newblock {\em Proceedings of the VLDB Endowment}, 3(1):48--57, 2010.
\newblock \href {https://doi.org/10.14778/1920841.1920853}
  {\path{doi:10.14778/1920841.1920853}}.

\bibitem{SPM}
Timothy~A. Davis and Yifan Hu.
\newblock The university of florida sparse matrix collection.
\newblock {\em ACM Transactions on Mathematical Software}, 38(1):1:1--1:25, 11
  2011.

\bibitem{ZOLTAN}
Karen~D. Devine, Erik~G. Boman, Robert~T. Heaphy, Rob~H. Bisseling, and
  Ümit~V. Catalyurek.
\newblock Parallel hypergraph partitioning for scientific computing.
\newblock In {\em IEEE Transactions on Parallel and Distributed Systems}, pages
  10--pp. IEEE, 2006.

\bibitem{PERFORMANCE-PROFILES}
Elizabeth~D. Dolan and Jorge~J. Mor{\'{e}}.
\newblock Benchmarking optimization software with performance profiles.
\newblock {\em Mathematical Programming}, 91(2):201--213, 2002.

\bibitem{FM}
Charles~M. Fiduccia and Robert~M. Mattheyses.
\newblock A linear-time heuristic for improving network partitions.
\newblock In {\em 19th Conference on Design Automation (DAC)}, pages 175--181,
  1982.

\bibitem{KAHYPAR-HFC}
Lars Gottesb{\"u}ren, Michael Hamann, Sebastian Schlag, and Dorothea Wagner.
\newblock Advanced flow-based multilevel hypergraph partitioning.
\newblock {\em 18th International Symposium on Experimental Algorithms (SEA)},
  2020.

\bibitem{REBAHFC}
Lars Gottesb{\"u}ren, Michael Hamann, and Dorothea Wagner.
\newblock {Evaluation of a Flow-Based Hypergraph Bipartitioning Algorithm}.
\newblock In {\em 27th European Symposium on Algorithms (ESA)}, pages
  52:1--52:17, 2019.
\newblock \href {https://doi.org/10.4230/LIPIcs.ESA.2019.52}
  {\path{doi:10.4230/LIPIcs.ESA.2019.52}}.

\bibitem{MT-KAHYPAR}
Lars Gottesbüren, Tobias Heuer, Peter Sanders, and Sebastian Schlag.
\newblock Scalable shared-memory hypergraph partitioning.
\newblock In {\em 23st Workshop on Algorithm Engineering \& Experiments
  (ALENEX)}. SIAM, 01 2021.

\bibitem{SLM}
Michael Hamann, Ben Strasser, Dorothea Wagner, and Tim Zeitz.
\newblock Distributed graph clustering using modularity and map equation.
\newblock In {\em European Conference on Parallel Processing (Euro-Par)}, pages
  688--702, 2018.
\newblock \href {https://doi.org/10.1007/978-3-319-96983-1\_49}
  {\path{doi:10.1007/978-3-319-96983-1\_49}}.

\bibitem{HeintzC14}
B.~Heintz and A.~Chandra.
\newblock Beyond graphs: Toward scalable hypergraph analysis systems.
\newblock {\em ACM {SIGMETRICS} Performance Evaluation Review}, 41(4):94--97,
  2014.

\bibitem{hg-processing-framework-MESH}
Benjamin Heintz, Rankyung Hong, Shivangi Singh, Gaurav Khandelwal, Corey
  Tesdahl, and Abhishek Chandra.
\newblock {MESH:} {A} flexible distributed hypergraph processing system.
\newblock {\em CoRR}, abs/1904.00549, 2019.
\newblock URL: \url{http://arxiv.org/abs/1904.00549}.

\bibitem{KAHYPAR-MF}
Tobias Heuer, Peter Sanders, and Sebastian Schlag.
\newblock Network flow-based refinement for multilevel hypergraph partitioning.
\newblock {\em {ACM} Journal of Experimental Algorithmics (JEA)},
  24(1):2.3:1--2.3:36, 09 2019.

\bibitem{KAHYPAR-CA}
Tobias Heuer and Sebastian Schlag.
\newblock Improving coarsening schemes for hypergraph partitioning by
  exploiting community structure.
\newblock In {\em 16th International Symposium on Experimental Algorithms
  (SEA)}, pages 21:1--21:19. Schloss Dagstuhl -- Leibniz-Zentrum f{\"u}r
  Informatik, 06 2017.

\bibitem{hg-processing-hyperx}
Wenkai Jiang, Jianzhong Qi, Jeffrey~Xu Yu, Jin Huang, and Rui Zhang.
\newblock Hyperx: {A} scalable hypergraph framework.
\newblock {\em {IEEE} Transactions on Knowledge and Data Engineering},
  31(5):909--922, 2019.
\newblock \href {https://doi.org/10.1109/TKDE.2018.2848257}
  {\path{doi:10.1109/TKDE.2018.2848257}}.

\bibitem{SHP}
Igor Kabiljo, Brian Karrer, Mayank Pundir, Sergey Pupyrev, Alon Shalita,
  Yaroslav Akhremtsev, and Alessandro Presta.
\newblock {Social Hash Partitioner: A Scalable Distributed Hypergraph
  Partitioner}.
\newblock volume~10, pages 1418--1429, 2017.
\newblock \href {https://doi.org/10.14778/3137628.3137650}
  {\path{doi:10.14778/3137628.3137650}}.

\bibitem{HMETIS}
George Karypis, Rajat Aggarwal, Vipin Kumar, and Shashi Shekhar.
\newblock Multilevel hypergraph partitioning: Applications in vlsi domain.
\newblock {\em IEEE Transactions on Very Large Scale Integration (VLSI)
  Systems}, 7(1):69--79, 1999.

\bibitem{HMETIS-K}
George Karypis and Vipin Kumar.
\newblock Multilevel $k$-way hypergraph partitioning.
\newblock Technical Report 98-036, University of Minnesota, 1998.

\bibitem{sword}
K~Ashwin Kumar, Abdul Quamar, Amol Deshpande, and Samir Khuller.
\newblock Sword: workload-aware data placement and replica selection for cloud
  data management systems.
\newblock {\em The VLDB Journal}, 23(6):845--870, 2014.

\bibitem{MT-METIS-REFINEMENT}
Dominique LaSalle and George Karypis.
\newblock A parallel hill-climbing refinement algorithm for graph partitioning.
\newblock In {\em 45th International Conference on Parallel Processing (ICPP)},
  pages 236--241. IEEE, 2016.

\bibitem{LENGAUER}
Thomas Lengauer.
\newblock {\em {Combinatorial Algorithms for Integrated Circuit Layout}}.
\newblock John Wiley \& Sons, Inc., 1990.

\bibitem{BIPART}
Sepideh Maleki, Udit Agarwal, Martin Burtscher, and Keshav Pingali.
\newblock {BiPart: A Parallel and Deterministic Multilevel Hypergraph
  Partitioner}.
\newblock In {\em Proceedings of the 26th ACM SIGPLAN Symposium on Principles
  and Practice of Parallel Programming}, pages 161--174, 2021.

\bibitem{MANN-PAPA14}
Zoltán~{\'{A}}. Mann and Pál~A. Papp.
\newblock Formula partitioning revisited.
\newblock In {\em 5th Pragmatics of {SAT} Workshop}, pages 41--56, 2014.

\bibitem{TheHypeIsOver}
Christian Mayer, Ruben Mayer, Sukanya Bhowmik, Lukas Epple, and Kurt Rothermel.
\newblock {HYPE: Massive Hypergraph Partitioning With Neighborhood Expansion}.
\newblock In {\em IEEE International Conference on Big Data}, pages 458--467.
  IEEE Computer Society, 2018.
\newblock \href {https://doi.org/10.1109/BigData.2018.8621968}
  {\path{doi:10.1109/BigData.2018.8621968}}.

\bibitem{PARHIP}
Henning Meyerhenke, Peter Sanders, and Christian Schulz.
\newblock Parallel graph partitioning for complex networks.
\newblock {\em IEEE Transactions on Parallel and Distributed Systems},
  28(9):2625--2638, 2017.

\bibitem{ADAPTIVE-STOP-RULE}
Vitaly Osipov and Peter Sanders.
\newblock n-level graph partitioning.
\newblock In {\em 18th European Symposium on Algorithms (ESA)}, pages 278--289.
  Springer, 2010.

\bibitem{PAPA-MARKOV}
David~A. Papa and Igor~L. Markov.
\newblock Hypergraph partitioning and clustering.
\newblock In {\em Handbook of Approximation Algorithms and Metaheuristics}.
  2007.

\bibitem{TBB}
Chuck Pheatt.
\newblock Intel threading building blocks.
\newblock {\em Journal of Computing Sciences in Colleges}, 23(4):298--298,
  2008.

\bibitem{LABEL_PROPAGATION}
Usha~Nandini Raghavan, R{\'e}ka Albert, and Soundar Kumara.
\newblock Near linear time algorithm to detect community structures in
  large-scale networks.
\newblock {\em Physical Review E}, 76(3):036106, 2007.

\bibitem{KAFFPA}
Peter Sanders and Christian Schulz.
\newblock {Engineering Multilevel Graph Partitioning Algorithms}.
\newblock In {\em 19th European Symposium on Algorithms (ESA)}, pages 469--480.
  Springer, 2011.

\bibitem{PHARD}
John~E. Savage and Markus~G. Wloka.
\newblock Parallelism in graph-partitioning.
\newblock {\em Journal of Parallel and Distributed Computing}, 13(3):257--272,
  1991.
\newblock \href {https://doi.org/10.1016/0743-7315(91)90074-J}
  {\path{doi:10.1016/0743-7315(91)90074-J}}.

\bibitem{KAHYPAR-DIS}
Sebastian Schlag.
\newblock High-quality hypergraph partitioning.
\newblock 2020.

\bibitem{KaHyPar-R}
Sebastian Schlag, Vitali Henne, Tobias Heuer, Henning Meyerhenke, Peter
  Sanders, and Christian Schulz.
\newblock $k$-way hypergraph partitioning via n-level recursive bisection.
\newblock In {\em 18th Workshop on Algorithm Engineering \& Experiments
  (ALENEX)}, pages 53--67. SIAM, 2016.

\bibitem{clay}
Marco Serafini, Rebecca Taft, Aaron~J Elmore, Andrew Pavlo, Ashraf Aboulnaga,
  and Michael Stonebraker.
\newblock Clay: Fine-grained adaptive partitioning for general database
  schemas.
\newblock {\em Proceedings of the VLDB Endowment}, 10(4):445--456, 2016.

\bibitem{SimonTeng97}
Horst~D. Simon and Shang{-}Hua Teng.
\newblock How good is recursive bisection?
\newblock {\em {SIAM} Journal of Scientific Computing}, 18(5):1436--1445, 1997.
\newblock \href {https://doi.org/10.1137/S1064827593255135}
  {\path{doi:10.1137/S1064827593255135}}.

\bibitem{PARALLEL-LOUVAIN}
Christian~L. Staudt and Henning Meyerhenke.
\newblock Engineering parallel algorithms for community detection in massive
  networks.
\newblock {\em IEEE Transactions on Parallel and Distributed Systems},
  27(1):171--184, 01 2016.

\bibitem{PARKWAY-2}
Aleksandar Trifunovic and William~J. Knottenbelt.
\newblock Parkway 2.0: A parallel multilevel hypergraph partitioning tool.
\newblock In {\em International Symposium on Computer and Information
  Sciences}, pages 789--800. Springer, 2004.

\bibitem{MONDRIAAN}
B.~Vastenhouw and R.~H. Bisseling.
\newblock {A Two-Dimensional Data Distribution Method for Parallel Sparse
  Matrix-Vector Multiplication}.
\newblock {\em {SIAM} Review}, 47(1):67--95, 2005.

\bibitem{DAC}
Natarajan Viswanathan, Charles~J. Alpert, Cliff C.~N. Sze, Zhuo Li, and
  Yaoguang Wei.
\newblock The dac 2012 routability-driven placement contest and benchmark
  suite.
\newblock In {\em 49th Conference on Design Automation (DAC)}, pages 774--782.
  ACM, 6 2012.

\bibitem{WILCOXON}
Frank Wilcoxon.
\newblock Individual comparisons by ranking methods.
\newblock In {\em Breakthroughs in Statistics}, pages 196--202. Springer, 1992.

\bibitem{hepart}
Wenyin Yang, Guojun Wang, Kim-Kwang~Raymond Choo, and Shuhong Chen.
\newblock Hepart: A balanced hypergraph partitioning algorithm for big data
  applications.
\newblock {\em Future Generation Computer Systems}, 83:250--268, 2018.

\end{thebibliography}

\clearpage

\appendix

\section{Pseudocode for Handling Contraction Dependencies}
\label{appendix:contract_deps}

\begin{algorithm2e}
	\KwIn{Vertex pair $(u,v)$ to contract}
	\SetEndCharOfAlgoLine{}
  \caption{Handling Contraction Dependencies}
  \label{algo:contract_deps}
	\FuncSty{lock($v$)} \;
	\If() {$\rep[v] \neq v$}{
		\FuncSty{unlock($v$)} and \Return \tcp*[r]{other thread contracts v}
	}
	\While(\label{pseudocode:comp_check:start_tree_walk}) {$\rep[u] \neq u$ and $\pending[u] = 0$}{
		$u \gets \rep[u]$ \;
		\If() {$u = v$}{
			\FuncSty{unlock($v$)} and \Return \tcp*[r]{cycle in $\contractionforest$}
		}
	}
	\If() {$v < u$}{
		\FuncSty{lock($u$)}
	}
	\Else {
		\FuncSty{unlock($v$)}, \FuncSty{lock($u$)}, \FuncSty{lock($v$)} \tcp*[r]{avoid deadlock}
		\If(){$\rep[v] \neq v$}{
			\FuncSty{unlock($v$)} and \Return \tcp*[r]{other thread contracts v}
		}
	}

	\If() {$\rep[u] = u$ or $\pending[u] > 0$} {
		$x \gets u$\;
		\While() {$\rep[x] \neq x$}{
			$x \gets \rep[x]$ \;
			\If(){$x = v$}{
				\FuncSty{unlock($v$)}, \FuncSty{unlock($u$)} and \Return \tcp*[r]{cycle in $\contractionforest$}
			}
		}
		$\rep[v] \gets u$	\tcp*[r]{suitable ancestor found}
		$\pending[u] \gets \pending[u] + 1$\;
		\FuncSty{unlock($u$)}, \FuncSty{unlock($v$)}\;
		\tcp{start performing contraction}
		\While(){$u \neq v$}{
			\If(\tcp*[f]{check with lock for $v$}){$\pending[v] = 0$}{
				\FuncSty{performContraction($u,v$)} \;
				$\pending[u] \gets \pending[u]-1$ \tcp*[r]{with lock for $u$}
				$v \gets u$ \;
				$u \gets \rep[u]$
			} \Else{
				\Return \tcp*[r]{transfer responsibility}
			}
		}

	}
	\Else() {
		\FuncSty{unlock($u$)} and goto Line~\ref{pseudocode:comp_check:start_tree_walk} \tcp*[r]{retry with different ancestor}
	}
\end{algorithm2e}

\clearpage

\section{Benchmark Set Statistics}
\label{appendix:benchmark_stats}

\begin{figure*}[!htb]
	\centering
	\vspace{-1.65cm}
	\externalizedfigure{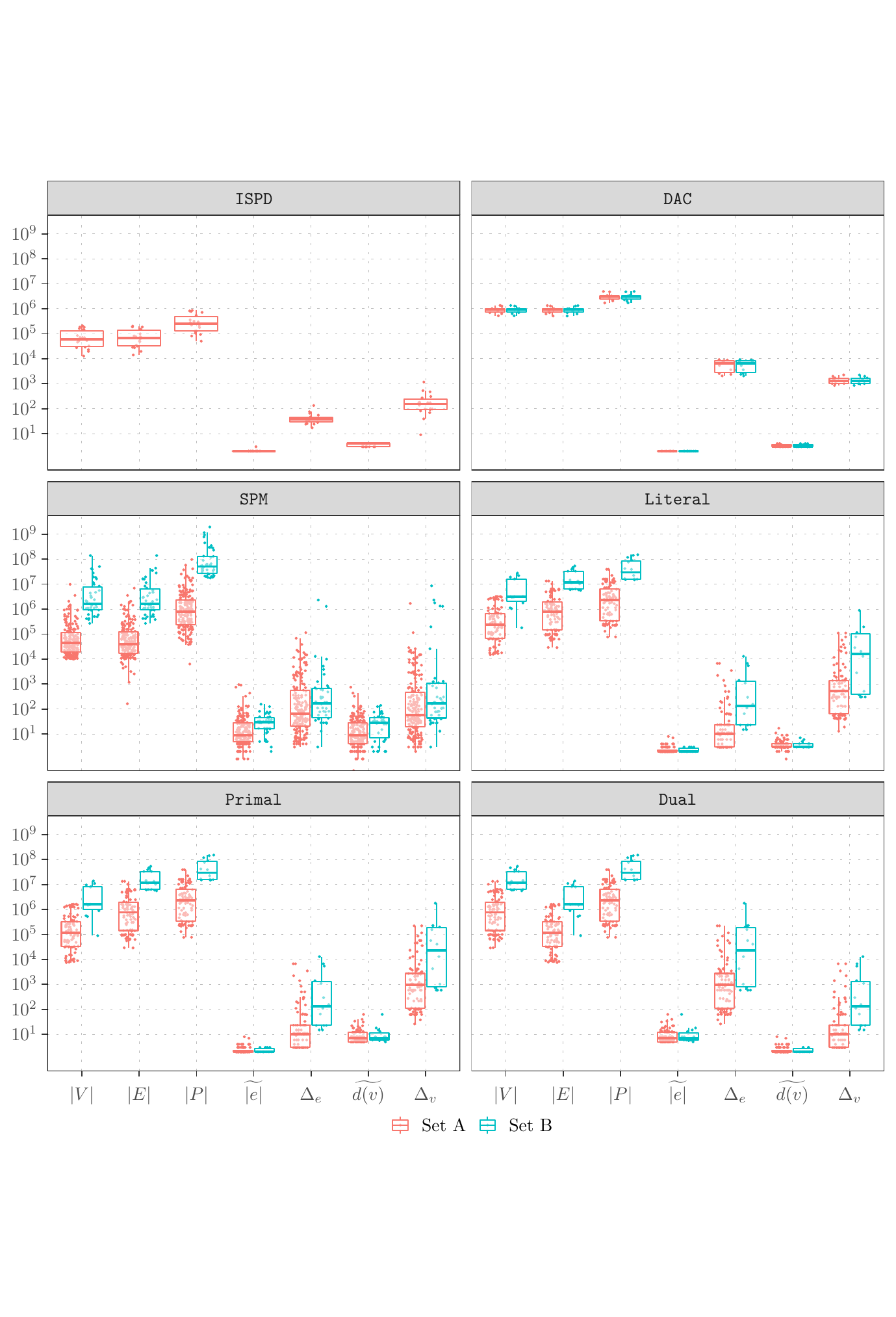}{experiments/benchmarks/benchmark_stats_per_type.tex} %
	\vspace{-3.25cm}
  \caption{Summary of different properties for our two benchmark sets and the different sources. It shows for each
           hypergraph (points), the number of vertices $|V|$, nets $|E|$ and pins $|P|$, as well as the median and maximum
           net size ($\medsize$ and $\maxsize{e}$ and vertex degree ($\meddeg$ and $\maxsize{v}$).}
	\label{fig:benchmark_set}
\end{figure*}

\end{document}